\documentclass[lettersize,journal]{IEEEtran}
\usepackage{amsmath,amsfonts}
\usepackage{esint}
\usepackage{array}
\usepackage{textcomp}
\usepackage{stfloats}
\usepackage{url}
\usepackage{verbatim}
\usepackage{graphicx}
\usepackage{cite}
\usepackage{tikz}
\usepackage{amssymb}
\usepackage{booktabs}
\usepackage{tabularx}
\usetikzlibrary{calc}
\usetikzlibrary{decorations}
\usepackage{pgfplots}
\pgfplotsset{compat=1.18}
\usepackage{algorithm}
\usepackage{algpseudocode}
\usepackage{multirow}
\usepackage[caption=false,font=footnotesize]{subfig}
\usepackage{xr}
\usepackage{booktabs}
\usepackage{makecell}
\usetikzlibrary{arrows.meta}

\newtheorem{proposition}{Proposition}

\newcolumntype{C}[1]{>{\centering\arraybackslash}m{#1}}

\newtheorem{lemma}{Lemma}

\begin{document}

\title{From Discrete Snapshots to Geometric Entity: Motion-Native Channel Tracking for THz Near-Field XL-MIMO}


\author{Yuheng Fan, Chuang Yang,~\IEEEmembership{Member,~IEEE}, Mugen Peng,~\IEEEmembership{Fellow,~IEEE}
\thanks{This work has been submitted to the IEEE for possible publication. Copyright may be transferred without notice, after which this version may no longer be accessible.}
\thanks{This work was supported by the Beijing Natural Science Foundation under grant L223007, in part supported by the Base Independent Scientific Research Project under Grant NST20250507, in part supported by BUPT Excellent Ph.D. Students Foundation CX20251017, in part supported by BUPT innovation and entrepreneurship support program  2025-YC-T001, and in part supported by Doctoral Student Program of the Young S$\&$T Talents Cultivation Project, CAST. \textit{(Corresponding author: Chuang Yang)}

Yuheng Fan, Chuang Yang, and Mugen Peng are with the State Key Laboratory of Networking and Switching Technology, Beijing University of Posts and Telecommunications, Beijing 100876, China (email: vanyh@bupt.edu.cn; chuangyang@bupt.edu.cn; pmg@bupt.edu.cn).}
}


\markboth{Author-prepared preprint}%
{Fan \MakeLowercase{\textit{et al.}}: Motion-Native Channel Tracking for THz Near-Field XL-MIMO}

\maketitle

\begin{abstract}
Channel tracking in mobile terahertz (THz) near-field extremely large-scale multiple-input multiple-output (XL-MIMO) systems is challenging because spherical-wave propagation couples angle and range, leading to rapid and nonlinear channel evolution under sparse pilot observations. Existing approaches typically estimate pilot-local channel or position snapshots before applying temporal tracking or smoothing, making them vulnerable to information loss during snapshot formation. To break this bottleneck, we propose a paradigm shift from tracking discrete snapshots to estimating a holistic geometric entity, namely, one shared continuous UE trajectory over the tracking window. Specifically, we parameterize the trajectory using a low-dimensional cubic B-spline representation and directly couple the per-pilot profile likelihoods through this shared trajectory before any pilot-local position estimate is finalized, thereby inducing continuous near-field channel evolution. We further characterize the efficient geometric information supplied by each pilot observation and its aggregation in the shared spline-coefficient space. Based on this formulation, we develop a regularized maximum a posteriori estimator and a quadratic-penalty lifted solver that separates parallel pilot-local geometric updates from a structured global spline update. The resulting global system is banded and, for a fixed spline degree, admits linear-complexity factorization, enabling an efficient structured implementation. Simulations demonstrate consistently improved tracking accuracy over representative temporally refined baselines in the tested medium-to-high-SNR regime, together with a favorable accuracy–runtime trade-off. These results highlight the benefit of coupling sparse pilot observations through a shared continuous trajectory rather than temporally processing independently estimated snapshots.
\end{abstract}


\section{Introduction}
\IEEEPARstart{T}{erahertz} (THz) communications and extremely large-scale multiple-input multiple-output (XL-MIMO) are widely regarded as the twin pillars of high-capacity wireless systems for the upcoming 6G era \cite{wang2023road}. By leveraging the enormous bandwidth at THz frequencies \cite{xue2024survey} and the high beamforming gain of XL-MIMO \cite{lu2024tutorial}, these systems can theoretically achieve Tbps-level data rates. However, the deployment of physically large arrays introduces severe near-field propagation effects, where the wavefronts impinging on the array are spherical rather than planar \cite{liu2023near}. Unlike conventional far-field channels that depend solely on angles, the near-field THz channel depends on both the user's angle and range. This added dimensionality makes instantaneous channel state information (CSI) sensitive to user movement and increases the demand for frequent pilot support in dynamic scenarios \cite{akyildiz2022terahertz}.

\subsection{Related Work and Limitations}

Existing near-field channel estimation methods can be broadly classified into sparsity-aware, parametric, and model-driven approaches
\cite{yang2025near}.
Sparsity-aware methods exploit structured channel representations in an
appropriate transform domain. Representative examples include on-grid
polar-domain orthogonal matching pursuit (OMP) and off-grid P-SIGW
\cite{cui2022channel}, near-field sparse representation
\cite{zhang2023near}, subarray-based hybrid-field estimation
\cite{sun2023subarray}, and Bayesian sparse reconstruction
\cite{pisharody2024near,zhang2026novel,zhu2025adaptive}.
Parametric methods instead estimate a compact set of physical parameters
directly in the continuous domain, including joint maximum-likelihood
(ML) estimation \cite{wang2024near}, tensor-based
methods \cite{hou2026tensor,liu2025tensor}, and sensing- or motion-assisted
estimation \cite{liu2025sensing,wang2024near,meng2025near}. Model-driven methods further
combine structured physical inference with trainable modules. Recent
hybrid-field examples include learned-sparsifying-matrix Bayesian estimation,
dual-attention deep-unfolded Bayesian learning, and model-driven deep learning
\cite{li2025hybridnearfar,li2026widebandhybrid,li2025modeldriven}. 

These methods provide effective slot-wise channel or physical-parameter
estimates and naturally serve as the front end of dynamic channel tracking.
Existing tracking
methods typically follow an estimate-then-track architecture. A static
estimator first recovers channel states or physical parameters at individual
pilot instants, after which temporal priors are used to propagate or smooth
these estimates. Deterministic methods track physical parameters such as
angle of arrival (AoA), delay, and Doppler through electromagnetic or
kinematic models, using techniques such as extended Kalman filtering
\cite{zhao2025near,long2025channel}. Rauch--Tung--Striebel (RTS) smoothing
\cite{zhang2026near} provides a state-space tracking reference. Statistical methods instead describe channel
evolution through Markov-type models and track latent sparse states
\cite{fan2025hybrid,fan2025dynamic}.  Temporal
consistency can also be introduced without a recursive state-space model.
For example, penalized B-spline smoothing
\cite{eilers1996flexible} fits a continuous trajectory to positions
estimated independently at the pilot instants. Although this approach
produces a continuous output, the pilot observations have already been
compressed into separate snapshot estimates before the spline is fitted.
The temporal model therefore refines the estimated snapshots rather than
participating directly in the observation-level inference.

\begin{table*}[t]
	\centering
	\caption{Positioning of representative near-field channel estimation and tracking paradigms.}
	\label{tab:paradigm_comparison}
	\footnotesize
	\setlength{\tabcolsep}{2pt}
	\renewcommand{\arraystretch}{1.15}
	\begin{tabularx}{\textwidth}{@{}
			>{\raggedright\arraybackslash}p{0.24\textwidth}
			>{\raggedright\arraybackslash}p{0.20\textwidth}
			>{\raggedright\arraybackslash}p{0.17\textwidth}
			>{\raggedright\arraybackslash}p{0.17\textwidth}
			>{\raggedright\arraybackslash}X@{}}
		\toprule
		\textbf{Representative direction}
		&
		\textbf{Inference target}
		&
		\textbf{Temporal coupling}
		&
		\textbf{Observation interface}
		&
		\textbf{Inference principle}
		\\
		\midrule
		Polar-domain estimation~\cite{cui2022channel}
		&
		Snapshot CSI
		&
		None
		&
		Pilot-local observations
		&
		Grid/off-grid sparse recovery
		\\
		Parametric estimation~\cite{wang2024near}
		&
		Physical parameters
		&
		None
		&
		Pilot-local observations
		&
		Slot-wise likelihood fitting
		\\
		Deterministic tracking~\cite{zhao2025near,zhang2026near}
		&
		Discrete physical states
		&
		Kinematic evolution
		&
		Snapshot sequence
		&
		Estimate--then--track
		\\
		Statistical tracking~\cite{fan2025hybrid,fan2025dynamic}
		&
		Latent channel states
		&
		Statistical evolution
		&
		Snapshot sequence
		&
		Estimate--then--track
		\\
		B-spline post-processing~\cite{eilers1996flexible}
		&
		Fitted continuous trajectory
		&
		Spline smoothing
		&
		Position snapshots
		&
		Estimate--then--smooth
		\\
		\textbf{Proposed}
		&
		\textbf{continuous channel}
		&
		\textbf{Window-level coupling}
		&
		\textbf{Joint likelihoods}
		&
		\textbf{Direct inference}
		\\
		\bottomrule
		\multicolumn{5}{p{0.96\textwidth}}{
			\footnotesize
			\textit{Note:} Representative formulation-level characteristics;
			the table does not provide a numerical performance ranking.
		}
	\end{tabularx}
\end{table*}

Despite their different estimators and temporal models, these approaches first compress each pilot observation into a local state during the estimation stage and impose temporal coupling afterward during the tracking stage. Under parameter-independent processing, observation-level information discarded during this compression cannot be recovered by subsequent temporal processing according to the data processing inequality
\cite{cover2006elements,merhav2012data}. This motivates coupling the pilot observations directly through the shared trajectory itself.

It is also useful to distinguish the above tracking problem from sensing-assisted and integrated sensing and communication (ISAC) formulations~\cite{liu2025sensing,zhao2025near}. From the perspective relevant to this paper, ISAC primarily addresses how to enrich the available geometric observations through sensing and communication resources, whereas the present work addresses how sparse communication-pilot observations are jointly inferred over time. The two directions are therefore complementary.

\subsection{Motivation and Contributions}

\begin{figure*}[t]
	\centering
	\includegraphics[width=\textwidth,keepaspectratio]{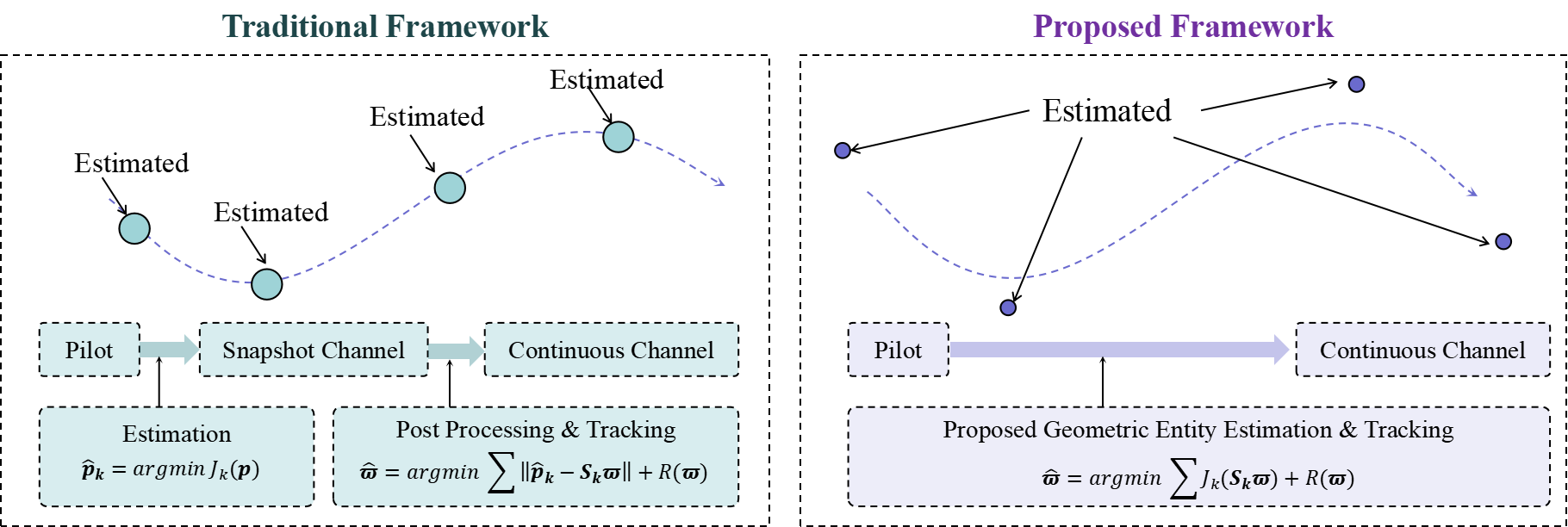}
	\caption{From snapshot-first estimation and temporal post-processing to
		motion-native continuous near-field channel tracking.}
	\label{Fig_01}
\end{figure*}

Motivated by this distinction, we incorporate estimation and temporal tracking directly into
the observation-level inference rather than imposing it on independently estimated snapshots. In this way, the continuous geometric process becomes the primary inference object, rather than a post-processed sequence of channel or position snapshots as illustrated in
Fig.~\ref{Fig_01}. The main contributions of this paper are summarized as follows:
\begin{itemize}
	
	\item\textbf{Formulate a motion-native continuous channel tracking problem:} We reformulate sparse-pilot near-field channel tracking as direct inference of a shared continuous UE trajectory rather than independent pilot-local snapshots. A cubic B-spline representation is introduced to parameterize the trajectory, which couples all pilot observations through a common geometric entity before any pilot-local state is finalized. Unlike estimate-then-smooth pipelines, the temporal model acts on the per-pilot profile likelihoods before any position estimate is finalized, retaining the shared trajectory as the primary inference variable.
	
	\item\textbf{Characterize the information properties and derive the lifted optimization structure:} After analytically eliminating the array-common complex coefficients, we characterize the observation-supported identifiability conditions of the shared spline trajectory and quantify the information loss caused by snapshot-first processing. Based on these properties, we further develop a finite quadratic-penalty lifted formulation.
	
	\item \textbf{Develop a structure-exploiting local–global solver}: By leveraging the parallel separability of pilot-local updates in the lifted formulation and the compact support of B-spline bases, the proposed solver enables parallel local refinement and banded global updates, achieving linear-complexity factorization with respect to the number of spline coefficients for a fixed spline degree.
\end{itemize}
Extensive simulations further validate the resulting tracking gain, operating scope, optimization behavior, and tested accuracy–runtime trade-off.

\subsection{Organization and Notation}
The remainder of this paper is organized as follows. Section II introduces the system model and formulates the continuous channel-tracking problem. Section III develops the regularized MAP formulation, analyzes observation-supported identifiability, and introduces the lifted penalty surrogate. Section IV presents the alternating solver and its structured implementation. Section V reports the simulation results, and Section VI concludes the paper.

\textit{Notation}: In this paper, bold uppercase and lowercase letters stand for matrices and vectors, respectively. Unless otherwise stated, all vectors are column vectors. Superscripts $(\cdot)^{T}$, $(\cdot)^{H}$, and $(\cdot)^{-1}$ denote the transpose, conjugate transpose, and inverse operations. The vectorization of a matrix is denoted as $\mathrm{vec}(\cdot)$, whose reverse process is $\mathrm{devec}(\cdot)$. The estimation of a variable $\mathbf{X}$ is denoted as $\hat{\mathbf{X}}$. The symbol $\otimes$ stands for the Kronecker product and $\mathbf I_x$ is the $x\times x$ identity matrix.

\section{System Model}

In this system, we assume different user-equipment (UE) terminals are assigned orthogonal pilots across
the $N$-slot time resource block.  Here, we consider a synchronized and calibrated system with dominant trackable LoS path and a fixed receiver-known combiner $\mathbf{W}$. Accordingly, we focus
on a single typical UE served by a multi-antenna base station (BS) equipped
with $M$ antennas connected to $N_{\rm RF}$ radio-frequency (RF) chains. The horizontal axis of the array is aligned with global x-axis in the two-dimensional Cartesian coordinate system and we assume that the UE lies in the known service half-plane to remove the standard mirror ambiguity of a ULA in the following sections. 

\begin{figure}[!t]
    \centering
    \includegraphics[width=\columnwidth,keepaspectratio]
    {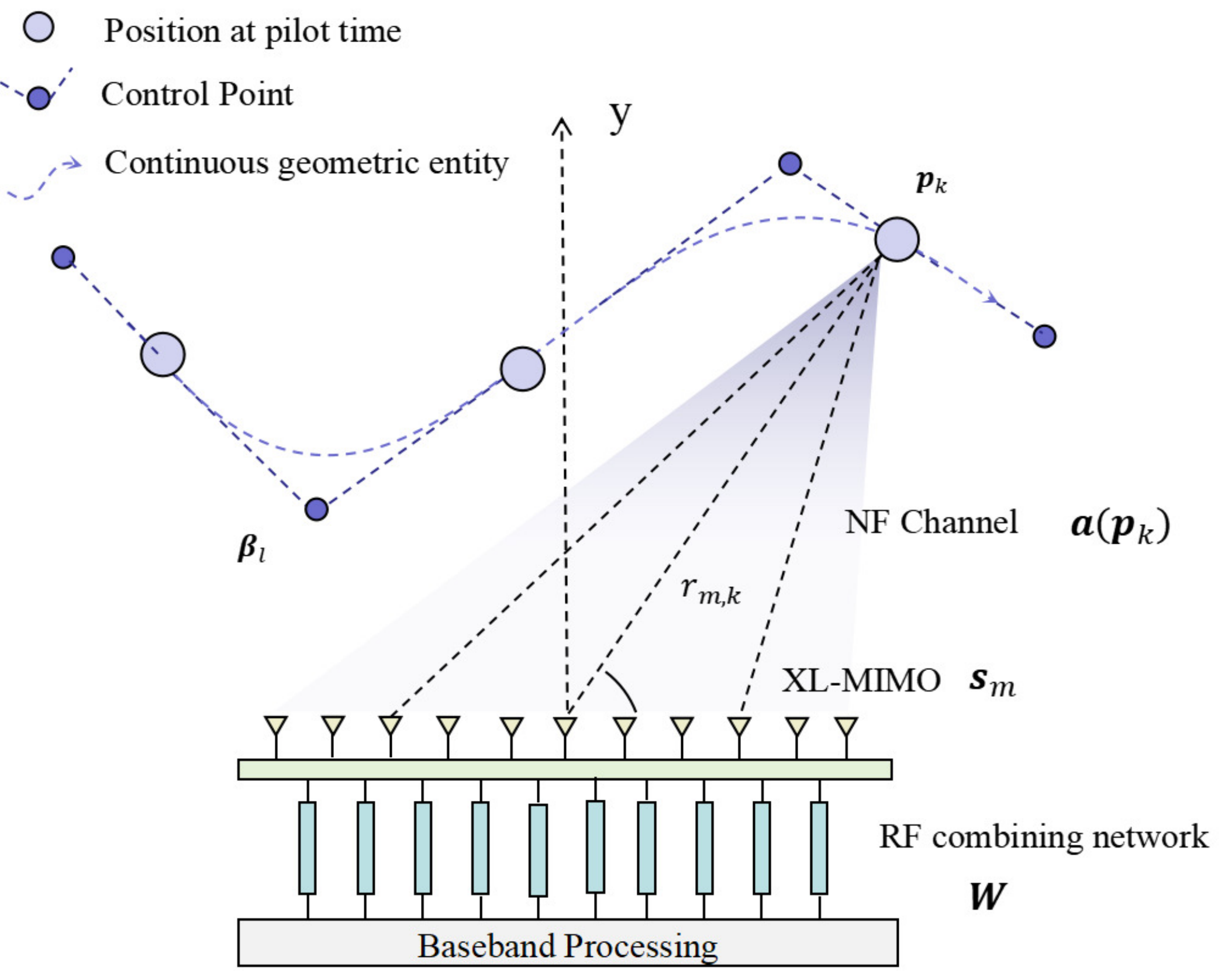}\\
    \vspace{0.6em}
    \includegraphics[width=\columnwidth]{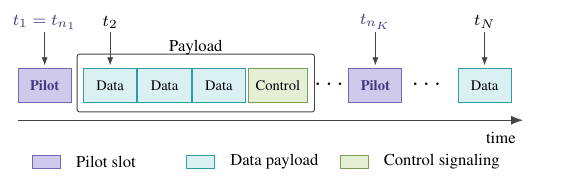}
    \caption{Pilot-supported time grid with pilot-bearing and pilot-free physical slots.}
    \label{Fig_02_System}
\end{figure}

In practical communication systems, the full observation window can be
discretized as $\mathcal{T} = \left\{t_{1},t_{2},\cdots,t_{N}\right\}$ and
observations are only available at a subset of instants, denoted by
$\mathcal{O}=\left\{t_{n_{1}},t_{n_{2}},\cdots,t_{n_{K}}\right\},~~~~K\ll N$.
The $K$ pilot-bearing slots provide the observations used for tracking,
whereas the remaining $N-K$ slots may carry payload data or control signaling.
Hence, $K/N$ denotes temporal pilot support rather than complete
time--frequency overhead, as illustrated in Fig.~\ref{Fig_02_System}. In this paper, the geometric entity refers to the complete UE trajectory over one tracking window. Motion-native means that this trajectory enters the joint pilot likelihood before any pilot-local channel or position estimate is finalized. 
	
\subsection{THz Near-Field XL-MIMO Signaling Model}

For a half-wavelength-spaced ULA, the position of $m$th antenna element is denoted as $\mathbf s_m =\left[\left(m-(M+1)/2\right)\frac{\lambda}{2},\,0\right]^{\mathsf T},$ where $\lambda$ is the wavelength at carrier frequency $f_c$. The UE position at the $k$th pilot instant is denoted by
$\mathbf p_k=\mathbf p(t_{n_k}) =[p_{x,k},p_{y,k}]^{\mathsf T}$. Therefore, the
propagation vector to the $m$th antenna element is $	\mathbf{r}_{m,k}=\mathbf p_{k}-\mathbf s_m$ and the distance ${d}_{m}(\mathbf p) = |\mathbf p-\mathbf s_m|_{2}$. Based on these element-dependent propagation distances, the spherical-wave
spatial response $\mathbf a(\mathbf p)\in\mathbb C^M$ in the near-field region is modeled as
\begin{equation}
	[\mathbf a(\mathbf p)]_m
	=
	\exp\!\left(
	-\mathrm j\frac{2\pi f_c}{c}{d}_{m}(\mathbf p) 
	\right).
	\label{eq:array_response}
\end{equation}
Besides this position-dependent phase-only near field response, $\alpha(t)$ is an array-common complex coefficient that absorbs other slowly varying effects, including residual common phase, pathloss, and synchronization/calibration errors. Because $\alpha(t)$ usually varies much more slowly than the geometry-induced spatial response $\mathbf a(\mathbf p)$, in this paper we assume that the coherence time of $\alpha(t)$ exceeds the inter-pilot interval. Accordingly, we adopt the blockwise-constant model \(\alpha(t)=\alpha_k\) for \(t\in[t_{n_k},t_{n_{k+1}})\), \(k=1,\ldots,K-1\), with \(\alpha_K\) used over the final interval \([t_{n_K},t_N]\).

Thus, the dominant-path channel is modeled as $\mathbf h(t)=\alpha(t)\mathbf a\!\left(\mathbf p(t)\right), \mathbf \mathbf h_{k}= \mathbf h(t_{n_{k}})$ and the corresponding post-combining observations can be written as $\mathbf Y_k =\mathbf{W} \mathbf{h}_{k}\boldsymbol\gamma_k^{\mathsf T}+\mathbf{W}\mathbf{N}_{k}\in\mathbb C^{N_{\rm RF}\times T_{\rm p}}$, where $\boldsymbol\gamma_k$ is the pilot block and
$\mathbf N_k$ denotes the antenna-domain noise matrix. Equivalently, vectorizing the received $\mathbf Y_k$ gives
\begin{equation}
	\begin{aligned}
		\mathbf y_k=
		\alpha_k\boldsymbol\psi_k(\mathbf p_k)
		+\mathbf n_k,
	\end{aligned}
	\label{eq:receiver equation}
\end{equation}
where $N_{y}$ denotes the dimension of the vectorized pilot observation, $\mathbf y_k,\mathbf n_k,\boldsymbol\psi_k(\mathbf p_k)\in\mathbb{C}^{N_{y}}$. $\mathbf y_k$ is the received signal, 
$\boldsymbol\psi_k(\mathbf p)\triangleq\boldsymbol\gamma_k\otimes\mathbf W\mathbf a(\mathbf p)$  maps a trial UE position directly to the
corresponding noiseless post-combining pilot response, $\alpha_k$
accounts for the common complex scaling. Since the combining matrix
$\mathbf W\in\mathbb C^{N_{\rm RF}\times M}$ is fixed across the tracking
window, known at the receiver, and semiunitary $\mathbf n_{k} = vec(\mathbf{W}\mathbf{N}_{k})\sim
\mathcal{CN}(\mathbf 0,\sigma^2\mathbf I_{N_{y}})$. Consequently, the corresponding likelihood can be expressed using Euclidean
residuals with the common scaling factor $1/\sigma^2$.
\begin{table}[t]
	\centering
	\caption{Principal notation and dimensions.}
	\label{tab:principal_notation}
	\renewcommand{\arraystretch}{1.04}
	\footnotesize
	\setlength{\tabcolsep}{3pt}
	
	\begin{tabularx}{\columnwidth}{@{}p{0.50\columnwidth}X@{}}
		\toprule
		\textbf{Symbol and domain} & \textbf{Meaning} \\
		\midrule
		
		$\mathbf p_k(\boldsymbol\varpi)
		=\mathbf S_k\boldsymbol\varpi\in\mathbb R^2$
		&
		Trajectory position at $t_{n_k}$
		\\
		
		$\mathbf a(\mathbf p)\in\mathbb C^M$
		&
		Near-field channel response
		\\
		
		$\mathbf W\in\mathbb C^{N_{\rm RF}\times M}$
		&
		Fixed semiunitary combiner
		\\
		
		$\boldsymbol\psi_k(\mathbf p)
		\in\mathbb C^{N_{\rm y}}$
		&
		Parameterized sensing vector
		\\
		
		$\alpha_k\in\mathbb C$
		&
		Array-common nuisance coefficient
		\\
		
		$\boldsymbol\varpi\in\mathbb R^{2L}$
		&
		Stacked B-spline coefficients
		\\
		
		$\mathbf S_k\in\mathbb R^{2\times2L}$
		&
		Spline sampling matrix at $t_{n_k}$
		\\
		
		$\mathbf z_k\in\mathbb R^2$
		&
		Auxiliary local position variable
		\\
		
		$\mathbf D_x\in\mathbb R^{2(L-x)\times2L},
		\ x\in\{1,2\}$
		&
		$x$-order difference operator
		\\
		
		$\mathbf R\in\mathbb R^{2L\times2L}$
		&
		Total regularization matrix
		\\
		
		$\mathbf C_{\boldsymbol\varpi}
		\in\mathbb R^{2L\times2L}$
		&
		Global normal matrix
		\\
		
		$\boldsymbol\Gamma_{\boldsymbol\varpi}
		\in\mathbb R^{2L}$
		&
		Global right-hand side
		\\
		
		\bottomrule
	\end{tabularx}
\end{table}

\subsection{Continuous-Time B-Spline Parametrization}

Rather than treating positions as independent variables, we regard the continuous UE trajectory as
a motion-native geometric entity. We represent the continuous UE trajectory by a degree-$q$ B-spline with L two-dimensional control points $\boldsymbol\beta_\ell
=[\beta_{x,\ell},\beta_{y,\ell}]^{\mathsf T}\in\mathbb R^2$, stacked as
\begin{equation}
	\boldsymbol\varpi
	\triangleq
	\left[
	\boldsymbol\beta_0^{\mathsf T},
	\ldots,
	\boldsymbol\beta_{L-1}^{\mathsf T}
	\right]^{\mathsf T}
	\in\mathbb R^{2L},
	\label{eq:spline_coefficients}
\end{equation}
where $L\ll N$ controls the dimension of the trajectory representation. Using normalized time
$s(t)\triangleq(t-t_1)/(t_N-t_1)\in[0,1]$
and adopting a clamped open-uniform knot sequence,
$\{\tau_i\}_{i=0}^{L+q}$ can be given by
\begin{equation}
	\tau_i
	=
	\begin{cases}
		0, & 0\leq i\leq q,\\[1mm]
		\dfrac{i-q}{L-q}, & q+1\leq i\leq L-1,\\[2mm]
		1, & L\leq i\leq L+q,
	\end{cases}
	\label{eq:clamped_knot_sequence}
\end{equation}
The repeated endpoint knots clamp the trajectory at both ends. The basis functions follow the Cox–de Boor recursion \cite{bohm1981generating,de1978practical} as
\begin{equation}
	B_{\ell,q}(s)
	=
	\frac{s-\tau_\ell}
	{\tau_{\ell+q}-\tau_\ell}
	B_{\ell,q-1}(s)
	+
	\frac{\tau_{\ell+q+1}-s}
	{\tau_{\ell+q+1}-\tau_{\ell+1}}
	B_{\ell+1,q-1}(s).
	\label{eq:cox_de_boor}
\end{equation}
At degree zero,
$B_{\ell,0}(s)$ equals one over its associated half-open knot interval and
zero elsewhere, with the final interval including the right endpoint.

Collecting the $L$ basis functions evaluated at $s(t)$,
we define $\mathbf b(t)\triangleq\big[B_{0,q}(s(t)),\ldots,B_{L-1,q}(s(t))
\big]^{\mathsf T}\in\mathbb R^L$ and the corresponding position basis matrix $\mathbf S(t)\triangleq \mathbf b^{\mathsf T}(t)\otimes\mathbf I_2
\in\mathbb R^{2\times2L}.$ The continuous trajectory generated by the control points can therefore be
written equivalently as
\begin{equation}
	\mathbf p(t;\boldsymbol\varpi)
	=
	\sum_{\ell=0}^{L-1}
	B_{\ell,q}(s(t))\boldsymbol\beta_\ell
	=
	\mathbf S(t)\boldsymbol\varpi.
	\label{eq:continuous_position}
\end{equation}

At the $k$th pilot instant, we further define the sampling matrix $\mathbf S_k\triangleq\mathbf S(t_{n_k})$, thus $\mathbf p_k(\boldsymbol\varpi)=\mathbf S_k\boldsymbol\varpi$. Substituting $\mathbf p_k(\boldsymbol\varpi)$ into the signaling model makes
every pilot observation depend on the same global coefficient vector
$\boldsymbol\varpi$. The B-spline representation therefore reduces the
trajectory description from $2N$ grid-wise position variables to $2L$
coefficients, while introducing temporal coupling directly at the
observation level throughout the tracking window.

We use cubic splines $(q=3)$, the lowest degree providing $C^{2}$-continuous trajectories. Their compact support activates at most four consecutive basis functions, preserving the banded global structure while providing sufficient smoothness for the considered motion.

\subsection{Problem Formulation}

Combining the near-field observation model with the spline
parameterization, each pilot observation depends on the same global
trajectory coefficient vector $\boldsymbol\varpi$ through
$\mathbf p_k(\boldsymbol\varpi)=\mathbf S_k\boldsymbol\varpi$, while retaining
its own array-common nuisance coefficient $\alpha_k$. Our objective is to
estimate $\boldsymbol\varpi$ and thereby recover the continuous trajectory
$\mathbf p(t)$. For compactness, collect the nuisance coefficients as $\boldsymbol\alpha\triangleq[\alpha_1,\ldots,\alpha_K]^{\mathsf T}\in\mathbb C^K.$ Under the white Gaussian post-combining noise model, the negative
log-likelihood, up to an additive constant, is
\begin{equation}
	J_0(\boldsymbol\varpi,\boldsymbol\alpha)
	=
	\frac{1}{\sigma^2}
	\sum_{k=1}^{K}
	\left\|
	\mathbf y_k
	-
	\alpha_k
	\boldsymbol\psi_k
	\bigl(\mathbf S_k\boldsymbol\varpi\bigr)
	\right\|_2^2.
	\label{eq:J0_cost}
\end{equation}
The corresponding unregularized maximum-likelihood estimator is therefore
\begin{equation}
	\mathrm P_0:\quad
	\bigl\{
	\widehat{\boldsymbol\varpi},
	\widehat{\boldsymbol\alpha}
	\bigr\}
	=
	\underset{\boldsymbol\varpi,\boldsymbol\alpha}{\arg\min}\;
	J_0(\boldsymbol\varpi,\boldsymbol\alpha).
	\label{eq:Q1}
\end{equation}

Although the spline parameterization reduces the trajectory representation complexity
from $2N$ grid-wise position variables to $2L$ global coefficients,
$\mathrm P_0$ remains a coupled nonlinear estimation problem. 

\section{Spline-Based MAP Framework with a Lifted Penalty Surrogate}

In this section, we incorporate regularization to
obtain a stable MAP formulation and eliminate the array-common nuisance coefficients analytically through variable projection starting from $\mathrm P_0$. We then discuss the
local identifiability provided by the near-field observations and the complementary stabilizing role of regularization. Finally, we introduce a lifted optimization framework, separating nonlinear pilot-local fitting from global spline consistency.

\subsection{Regularized MAP and Variable Projection}

Because practical UE trajectories are typically smooth, we exploit temporal smoothness to stabilize $\mathrm P_0$ in the sparse-pilot regime. From a Bayesian perspective, this regularization corresponds to an intrinsic Gaussian
prior
\begin{equation}
	p(\boldsymbol\varpi)
	\propto
	\exp[-\mathcal R(\boldsymbol\varpi)],
\end{equation}
which favors slowly varying control-point sequences through first- and
second-order differences that penalize velocity- and acceleration-like variations, respectively. In addition, since $\boldsymbol\varpi$ is the primary variable in this paper, the smoothness prior is imposed directly on the spline coefficients rather than on independently sampled positions.

Define the corresponding scalar difference operators for a generic coefficient sequence
$\mathbf x=[x_0,\ldots,x_{L-1}]^{\mathsf T}$ as
$\boldsymbol\Delta_1\in\mathbb R^{(L-1)\times L}$ and
$\boldsymbol\Delta_2\in\mathbb R^{(L-2)\times L}$, where
\begin{equation}
	[\boldsymbol\Delta_1\mathbf x]_\ell
	=
	x_{\ell+1}-x_\ell,
	\qquad
	[\boldsymbol\Delta_2\mathbf x]_\ell
	=
	x_{\ell+2}-2x_{\ell+1}+x_\ell .
	\label{eq:scalar_differences}
\end{equation}
These scalar operators are then extended to the two-dimensional coefficient vector as
\begin{equation}
	\mathbf D_x
	\triangleq
	\boldsymbol\Delta_x\otimes\mathbf I_2
	\in\mathbb R^{2(L-x)\times2L},
	\qquad x\in\{1,2\}.
\end{equation}
Accordingly,
$\mathbf D_1\boldsymbol\varpi$ captures the first-order
variation between adjacent control points, whereas
$\mathbf D_2\boldsymbol\varpi$ captures the variation of their
successive increments. Therefore, the prior regularizer is $\mathcal R(\boldsymbol\varpi)=\frac{1}{2}\boldsymbol\varpi^{\mathsf T}
\mathbf R\boldsymbol\varpi$
 where
 \begin{equation}
 	\mathbf R\triangleq\lambda_1\mathbf D_1^{\mathsf T}\mathbf D_1+\lambda_2\mathbf D_2^{\mathsf T}\mathbf D_2\in\mathbb R^{2L\times2L},
 \end{equation}
where nonnegative weights $\lambda_1$ and $\lambda_2$ control the penalties on control variation. Their effective values depend on the coordinate units and the relative scaling of the likelihood and regularization terms, and are
therefore treated as tuning parameters. Combining this regularization with the data fidelity in \eqref{eq:J0_cost}
gives
\begin{equation}
J_{\rm MAP}(\boldsymbol\varpi,\boldsymbol\alpha)
=J_0(\boldsymbol\varpi,\boldsymbol\alpha)
+\mathcal{R}(\boldsymbol\varpi),
\label{eq:J-map}
\end{equation}
and the corresponding regularized MAP problem is
\begin{equation}
\mathrm P_1:\quad
\{\widehat{\boldsymbol\varpi},\widehat{\boldsymbol\alpha}\}
=\underset{\boldsymbol\varpi,\boldsymbol\alpha}{\arg\min}\;
J_{\rm MAP}(\boldsymbol\varpi,\boldsymbol\alpha).
\label{eq:Q1-map}
\end{equation}

Although the trajectory regularization stabilizes
$\mathrm P_1$, the global spline coefficients and the $K$
array-common nuisance coefficients remain coupled through the
nonlinear position-dependent sensing model. For a fixed
trajectory coefficient vector $\boldsymbol\varpi$, the nuisance
coefficient $\alpha_k$ enters the $k$th observation model
linearly through the induced pilot position
$\mathbf p_k=\mathbf S_k\boldsymbol\varpi$.
For any trial position \(\mathbf p\), \(\alpha_k\) can be eliminated analytically by variable projection as
\begin{equation}
	\alpha_k^\star(\boldsymbol\varpi)
	=
	\frac{
		\boldsymbol\psi_k^{\mathsf H}
		(\mathbf S_k\boldsymbol\varpi)
		\mathbf y_k
	}{
		\boldsymbol\psi_k^{\mathsf H}
		(\mathbf S_k\boldsymbol\varpi)
		\boldsymbol\psi_k
		(\mathbf S_k\boldsymbol\varpi)
	},
	\label{eq:weighted_gain_projection}
\end{equation}
provided that
$\boldsymbol\psi_k^{\mathsf H}
(\mathbf S_k\boldsymbol\varpi)
\boldsymbol\psi_k
(\mathbf S_k\boldsymbol\varpi)>0$. After estimating \(\widehat{\boldsymbol\varpi}\), the nuisance coefficients are recovered as \(\widehat{\alpha}_k=\alpha_k^\star(\widehat{\boldsymbol\varpi})\), and the continuous channel is reconstructed accordingly. Substituting this closed-form solution into the observation residual yields 
\begin{equation}
	J_k(\mathbf S_k\boldsymbol\varpi)
	\triangleq
	\frac{1}{\sigma^2}
	\left\|
	\mathbf y_k
	-
	\alpha_k^\star(\mathbf S_k\boldsymbol\varpi)
	\boldsymbol\psi_k(\mathbf S_k\boldsymbol\varpi)
	\right\|_2^2.
	\label{eq:local_reduced_cost}
\end{equation}
where $J_k(\mathbf p)$ measures the variance-normalized residual energy after the best array-common complex scaling has been removed at the trial position
$\mathbf p$.

Applying this concentration step at all pilot instants converts $\mathrm P_1$
into the equivalent reduced objective
\begin{equation}
	J_{\rm dir}(\boldsymbol\varpi)
	=
	\sum_{k=1}^{K}
	J_k\!\left(\mathbf S_k\boldsymbol\varpi\right)
	+
	\mathcal{R}(\boldsymbol\varpi),
	\label{eq:direct_reduced_objective}
\end{equation}
which leaves the global spline coefficient vector $\boldsymbol\varpi$ as the primary
inference variable. 

Variable projection leaves $\boldsymbol{\varpi}$ as the sole inference variable, but $J_{\rm dir}$ remains nonlinear because all pilots are coupled through the shared spline. We next characterize the observation-supported information of this representation before developing the solver.

\subsection{Observation-Level Information and Identifiability}
The shared trajectory representation reduces the unknown trajectory
from discrete position samples to a finite-dimensional coefficient
vector. However, this reduction alone does not guarantee that the
coefficient vector is supported by the available pilot observations.
Therefore, we characterize the efficient information of
$\boldsymbol\varpi$ after eliminating the array-common nuisance
coefficients, and distinguish observation-supported local identifiability from regularization-induced curvature. For the theoretical analysis below, all information quantities are evaluated at the true parameter values.

\subsubsection{Identifiability of the Shared Trajectory Representation}

For the $k$th pilot, the coefficient-space Jacobian for post-combining response vector is
$\mathbf Q_{\boldsymbol\varpi,k}\triangleq\nabla\boldsymbol\psi_k(\mathbf S_k\boldsymbol\varpi)$. With a coefficient perturbation $\delta\boldsymbol\varpi$, only the component of this response variation orthogonal to the nuisance direction induced by the unknown coefficient $\alpha_{k}$ carries efficient information about $\boldsymbol\varpi$. Thus, we project out this nuisance direction using
\begin{equation}
	\mathbf P_{\psi,k}^{\perp}
	\triangleq
	\mathbf I_{N_{\rm y}}
	-
	\frac{\boldsymbol\psi_k(\mathbf S_k\boldsymbol\varpi)
		(\boldsymbol\psi_k(\mathbf S_k\boldsymbol\varpi))^{\mathsf H}}
	{\|\boldsymbol\psi_k(\mathbf S_k\boldsymbol\varpi)\|_2^2}.
	\label{eq:gain_orthogonal_projector}
\end{equation}
The efficient information contributed directly to the global
coefficient vector by the $k$th raw pilot observation is
\begin{equation}
	\mathcal I_{\boldsymbol\varpi,k}^{\rm obs}
	\triangleq
	\frac{2|\alpha_k|^2}{\sigma^2}
	\operatorname{Re}
	\left\{
	\mathbf Q_{\boldsymbol\varpi,k}^{\mathsf H}
	\mathbf P_{\psi,k}^{\perp}
	\mathbf Q_{\boldsymbol\varpi,k}
	\right\}.
	\label{eq:single_pilot_spline_information}
\end{equation}
To expose how pilot-local geometric sensitivity is mapped into the shared coefficient space, we obtain 
$\mathbf Q_{\mathbf p,k}\triangleq\nabla\boldsymbol\psi_k(\mathbf p)$ by the chain rule $\mathbf Q_{\boldsymbol\varpi,k}
=\mathbf Q_{\mathbf p,k}\mathbf S_k$. Thus, $\mathcal I_{\mathbf p,k}^{\rm obs}$ captures the independent
pilot-local geometric information
\begin{equation}
	\mathcal I_{\mathbf p,k}^{\rm obs}
	\triangleq
	\frac{2|\alpha_k|^2}{\sigma^2}
	\operatorname{Re}
	\left\{
	\mathbf Q_{\mathbf p,k}^{\mathsf H}
	\mathbf P_{\psi,k}^{\perp}
	\mathbf Q_{\mathbf p,k}
	\right\}.
\end{equation}
while
$\mathbf S_k^{\mathsf T}\mathcal I_{\mathbf p,k}^{\rm obs}\mathbf S_k$
maps it into the coefficient space, allowing different pilots to jointly constrain the same coefficients.

More specifically, under conditionally independent pilot observations, these coefficient-space contributions add as
\begin{equation}
	\mathcal I_{\boldsymbol\varpi}^{\rm obs}
	\triangleq
	\sum_{k=1}^{K}
	\mathcal I_{\boldsymbol\varpi,k}^{\rm obs}.
	\label{eq:total_coefficient_information}
\end{equation}
The shared coefficient vector $\boldsymbol\varpi$ is
first-order locally identifiable from the raw pilot
observations if and only if
$\mathcal I_{\boldsymbol\varpi}^{\rm obs}\succ\mathbf0$. To expose how the pilot-local information jointly constrains
the shared coefficients, we define the sampling matrix
$\mathbf A_{\mathcal O}
\triangleq$\footnotesize{$
\left[
(\tiny
\sqrt{\mathcal I_{\mathbf p,1}^{\rm obs}}\mathbf S_1
)^{\mathsf T},
\ldots,
(
\sqrt{\mathcal I_{\mathbf p,K}^{\rm obs}}\mathbf S_K
)^{\mathsf T}
\right]^{\mathsf T}$ \normalsize}$
\in\mathbb R^{2K\times2L}$, \normalsize where
\footnotesize$\sqrt{\mathcal I_{\mathbf p,k}^{\rm obs}}$\normalsize  ~denotes its symmetric positive-semidefinite square root. For the adopted B-spline representation, further define
$\mathbf B_{\mathcal O}
\triangleq
[\mathbf b(t_{n_1}),\ldots,
\mathbf b(t_{n_K})]^{\mathsf T}
\in\mathbb R^{K\times L}$. Therefore, this yields the following proposition.

\begin{proposition}[Information-weighted spline identifiability]
	\label{prop:weighted_sampling_identifiability}
	The shared coefficient vector $\boldsymbol\varpi$ is
	observation-supported first-order locally identifiable according to
	\begin{equation}
		\begin{aligned}
			\mathcal I_{\boldsymbol\varpi}^{\rm obs}\succ\mathbf0
			&\Longleftrightarrow
			\operatorname{rank}(\mathbf A_{\mathcal O})=2L
			\\[-1mm]
			&\Longrightarrow
			\sum_{k=1}^{K}
			\operatorname{rank}
			\bigl(\mathcal I_{\mathbf p,k}^{\rm obs}\bigr)
			\ge 2L
			\Longrightarrow K\ge L ,
			\label{eq:weighted_sampling_rank}
		\end{aligned}
	\end{equation}
	whereas, if
	$\mathcal I_{\mathbf p,k}^{\rm obs}\succ\mathbf0$ for all $k$,
	the exact condition reduces to
	\begin{equation}
		\mathcal I_{\boldsymbol\varpi}^{\rm obs}\succ\mathbf0
		\quad\Longleftrightarrow\quad
		\operatorname{rank}(\mathbf B_{\mathcal O})=L .
		\label{eq:bspline_collocation_rank}
	\end{equation}
\end{proposition}
The proofs are given in Appendix~\ref{app:identifiability_proof}. Condition~\eqref{eq:weighted_sampling_rank} allows rank-deficient pilot-local information as long as the mapped directions in $\mathbf S_{k}$ are complementary. Hence, a $2L$-dimensional spline trajectory can remain identifiable for $L\leq K\leq N$, whereas independently parameterized position cannot. Under full-rank local information, identifiability further requires
$\operatorname{rank}(\mathbf B_{\mathcal O})=L$, so $K\ge L$
alone is insufficient unless the pilot instants cover all spline
coefficient directions.

These rank conditions concern observation-supported identifiability,
whereas regularization contributes additional prior-induced curvature:
\begin{equation}
	\mathbf G_{\boldsymbol\varpi}^{\rm pen}
	=
	\mathcal I_{\boldsymbol\varpi}^{\rm obs}+\mathbf R .
\end{equation}
Regularization does not add observation information but only supplies prior curvature along weakly constrained coefficient directions. Its role is to suppress excessive oscillations of the spline trajectory under sparse and noisy observations, rather than to provide additional observation information.

\subsubsection{Information Preservation under Snapshot Compression}

Proposition~\ref{prop:weighted_sampling_identifiability}
shows how pilot-local geometric information can become complementary
after being mapped into the shared spline-coefficient space. A remaining question is whether the same benefit can be retained by first compressing each pilot observation into an independent snapshot and imposing the trajectory model afterward.

Consider the parameter-independent snapshot-first processing chain
\begin{equation}
	\mathbf y
	\xrightarrow{\;\mathcal K_{\rm s}\;}
	\widehat{\boldsymbol\xi}
	\xrightarrow{\;\mathcal K_{\rm p}\;}
	\widehat{\boldsymbol\zeta},
	\label{eq:Markovian Chain}
\end{equation}
where $\widehat{\boldsymbol\xi}$ denotes the pilot-local snapshot
sequence and $\widehat{\boldsymbol\zeta}$ its temporally processed
output. Let $\mathcal I_{\mathbf p,k}^{\rm snap}$ denote the
pilot-local efficient information retained after snapshot formation,
and define the information losses
\begin{equation}
	\Delta\mathcal I_{\mathbf p,k}^{\mathcal K_{\rm s}}
	\triangleq
	\mathcal I_{\mathbf p,k}^{\rm obs}
	-
	\mathcal I_{\mathbf p,k}^{\rm snap},
	\qquad
	\Delta\mathcal I_{\boldsymbol\varpi}^{\mathcal K_{\rm p}}
	\triangleq
	\mathcal I_{\boldsymbol\varpi}^{\rm snap}
	-
	\mathcal I_{\boldsymbol\varpi}^{\rm post},
	\label{eq:app_local_snapshot_loss}
\end{equation}
together with
$\Delta\mathcal I_{\boldsymbol\varpi}^{\rm tot}
\triangleq
\mathcal I_{\boldsymbol\varpi}^{\rm obs}
-
\mathcal I_{\boldsymbol\varpi}^{\rm post}$.
The resulting loss satisfies the following proposition.

\begin{proposition}[Irrecoverable information loss under snapshot-first processing]
	\label{prop:information_monotonicity}
	For the adopted shared B-spline representation, the total
	trajectory-information loss under snapshot-first processing satisfies
	\begin{equation}
		\Delta\mathcal I_{\boldsymbol\varpi}^{\rm tot}
		=
		\sum_{k=1}^{K}
		\mathbf S_k^{\mathsf T}
		\Delta\mathcal I_{\mathbf p,k}^{\mathcal K_{\rm s}}
		\mathbf S_k
		+
		\Delta\mathcal I_{\boldsymbol\varpi}^{\mathcal K_{\rm p}}
		\succeq\mathbf0 .
		\label{eq:total_information_loss}
	\end{equation}
	Furthermore,
	\begin{equation}
		\Delta\mathcal I_{\boldsymbol\varpi}^{\rm tot}
		=\mathbf0
		\Longleftrightarrow
		\Delta\mathcal I_{\mathbf p,k}^{\mathcal K_{\rm s}}
		=\mathbf0,\ \forall k,
		\quad
		\Delta\mathcal I_{\boldsymbol\varpi}^{\mathcal K_{\rm p}}
		=\mathbf0 .
		\label{eq:zero_information_loss}
	\end{equation}
\end{proposition}
The proof is provided in
Appendix~\ref{app:information_relation}.

Together, Propositions~\ref{prop:weighted_sampling_identifiability}
and~\ref{prop:information_monotonicity} capture the two sides of the proposed motion-native inference. The shared spline aggregates complementary pilot information, whereas information lost before this aggregation cannot be recovered by subsequent temporal processing. This motivates retaining direct observation-level coupling in \(J_{\rm dir}\) rather than adopting a snapshot-then-smooth pipeline.

\subsection{Lifted Surrogate and Block Decomposition}

To solve the nonlinear coupling in \(J_{\rm dir}(\boldsymbol\varpi)\), we introduce pilot-local auxiliary variables \(\mathbf z_k\in\mathbb R^2\) and promote local--global consistency through the finite quadratic penalty \(Z_k(\mathbf z_k,\boldsymbol\varpi)=\|\mathbf z_k-\mathbf S_k\boldsymbol\varpi\|_2^2\). The lifting introduces auxiliary optimization variables without changing the physical inference target $\boldsymbol\varpi$.

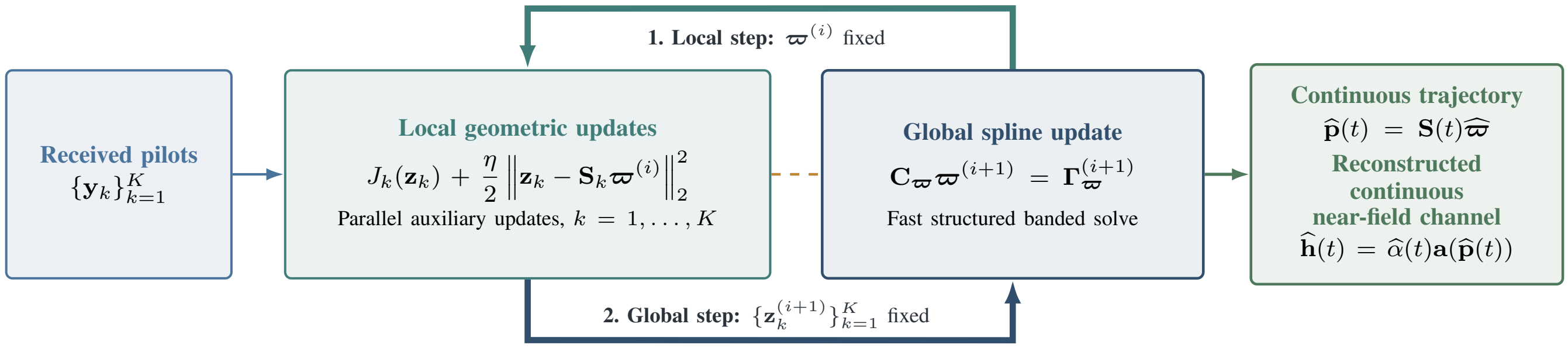
\begin{figure*}[!t]
	\centering
	
	\definecolor{figNavy}{RGB}{49,79,109}
	\definecolor{figBlue}{RGB}{74,116,156}
	\definecolor{figTeal}{RGB}{65,125,120}
	\definecolor{figAmber}{RGB}{191,135,52}
	\definecolor{figGreen}{RGB}{78,121,91}
	\definecolor{figInk}{RGB}{42,49,56}
	
	\begin{tikzpicture}[
		x=1cm,
		y=1cm,
		scale=0.98,
		transform shape,
		font=\footnotesize,
		stagebox/.style={
			rounded corners=2.2pt,
			line width=0.55pt
		},
		stageheader/.style={
			rounded corners=1.5pt,
			inner xsep=5pt,
			inner ysep=2.2pt,
			font=\scriptsize\bfseries,
			text=white,
			align=center
		},
		card/.style={
			rounded corners=2pt,
			line width=0.75pt,
			align=center,
			inner xsep=3mm,
			inner ysep=2.5mm,
			minimum height=24mm
		},		
		inputcard/.style={
			card,
			draw=figBlue,
			fill=figBlue!10,
			text width=20mm
		},
		localcard/.style={
			card,
			draw=figTeal,
			fill=figTeal!10,
			text width=50mm
		},
		globalcard/.style={
			card,
			draw=figNavy,
			fill=figNavy!9,
			line width=0.85pt,
			text width=38mm
		},		
		outputcard/.style={
			card,
			draw=figGreen,
			fill=figGreen!10,
			line width=0.85pt,
			text width=30mm
		},		
		flowlabel/.style={
			fill=white,
			rounded corners=1.5pt,
			inner xsep=3pt,
			inner ysep=1.5pt,
			font=\scriptsize,
			text=figInk,
			align=center
		},
		localsteplabel/.style={
			flowlabel,
			draw=none,
			fill=none
		},
		globalsteplabel/.style={
			flowlabel,
			draw=none,
			fill=none
		},
		dataflow/.style={
			-{Latex[length=2.0mm,width=1.4mm]},
			draw=figBlue,
			line width=0.8pt
		},
		consistencyflow/.style={
			draw=figAmber,
			dashed,
			line width=0.85pt
		},		
		outputflow/.style={
			-{Latex[length=2.0mm,width=1.4mm]},
			draw=figGreen,
			line width=1.0pt
		},
		localflow/.style={
			-{Latex[length=2.7mm,width=1.9mm]},
			draw=figTeal,
			line width=2pt
		},
		globalflow/.style={
			-{Latex[length=2.7mm,width=1.9mm]},
			draw=figNavy,
			line width=2pt
		}
		]
		\node[inputcard] (input) at (1.18,0) {
			\textcolor{figBlue}{\textbf{Received pilots}}
			\\[0.5mm]
			$\{\mathbf y_k\}_{k=1}^{K}$
		};
		\node[localcard](local)at(5.90,0){
			\textcolor{figTeal}{\textbf{Local geometric updates}}
			\\[1.2mm]
			$\displaystyle
			J_k(\mathbf z_k)
			+
			\frac{\eta}{2}
			\left\|
			\mathbf z_k-\mathbf S_k\boldsymbol\varpi^{(i)}
			\right\|_2^2$
			\\[1.0mm]
			{\scriptsize Parallel auxiliary updates, $k=1,\ldots,K$}
		};
		\node[globalcard] (global) at (11.5,0) {
			\textcolor{figNavy}{\textbf{Global spline update}}
			\\[1.5mm]
			$\displaystyle
			\mathbf C_{\boldsymbol\varpi}
			\boldsymbol\varpi^{(i+1)}
			=
			\boldsymbol\Gamma_{\boldsymbol\varpi}^{(i+1)}$
			\\[1.5mm]
			{\scriptsize Fast structured banded solve}
		};
		
		\node[outputcard] (output) at (16.05,0) {
			\textcolor{figGreen}{\textbf{Continuous trajectory}}
			\\[0.7mm]
			$\widehat{\mathbf p}(t)
			=
			\mathbf S(t)\widehat{\boldsymbol\varpi}$
			\\[0.9mm]
			\textcolor{figGreen}{\textbf{Reconstructed continuous}}
			\\[-0.2mm]
			\textcolor{figGreen}{\textbf{near-field channel}}
			\\[0.5mm]
			$\widehat{\mathbf h}(t)
			=
			\widehat{\alpha}(t)
			\mathbf a\!\left(\widehat{\mathbf p}(t)\right)$
		};
		
		
		\draw[dataflow]
		(input.east) -- (local.west);
		\draw[consistencyflow]
		(local.east) -- (global.west);
		\draw[outputflow]
		(global.east) -- (output.west);
		
		
		\coordinate (localStepFrom)
		at ([yshift=7mm]global.north);
		\coordinate (localStepTo)
		at ([yshift=7mm]local.north);
		\draw[localflow]
		(global.north)
		--
		(localStepFrom)
		--
		node[
		midway,
		below=2.5pt,
		localsteplabel
		] {
			\textbf{1. Local step:}\enspace
			$\boldsymbol\varpi^{(i)}$ fixed
		}
		(localStepTo)
		--
		(local.north);
		
		\coordinate (globalStepFrom)
		at ([yshift=-7mm]local.south);
		\coordinate (globalStepTo)
		at ([yshift=-7mm]global.south);
		\draw[globalflow]
		(local.south)
		--
		(globalStepFrom)
		--
		node[
		midway,
		above=1pt,
		globalsteplabel
		] {
			\textbf{2. Global step:}\enspace
			$\{\mathbf z_k^{(i+1)}\}_{k=1}^{K}$ fixed
		}
		(globalStepTo)
		--
		(global.south);
	\end{tikzpicture}
	\caption{Lifted local--global decomposition of the motion-native estimator.}
	
	\label{fig:lifted_penalty_framework}
\end{figure*}

Consequently, the
global continuous trajectory remains the primary inference object, while the
auxiliary variables merely expose the pilot-wise geometry needed for
efficient local updates. The resulting finite quadratic-penalty objective is
\begin{equation}
\begin{aligned}
J_\eta(\boldsymbol\varpi,\{\mathbf z_k\})
=\sum_{k=1}^{K}{\left[J_k(\mathbf z_k)+\frac{\eta}{2}
\mathcal{Z}_{k}(\mathbf{z}_{k},\boldsymbol{\varpi})\right]}	+\mathcal{R}(\boldsymbol\varpi),
\end{aligned}
\label{eq:lifted cost function}
\end{equation}
so that $\{\widehat{\boldsymbol\varpi},
\{\widehat{\mathbf z}_k\}_{k=1}^{K}\}
={\arg\min} J_\eta(\boldsymbol\varpi,\{\mathbf z_k\})$. The penalty encourages but does not impose exact equality for finite $\eta$, which naturally leads to a block-coordinate strategy. By
alternating between the $\{\mathbf z_k\}_{k=1}^{K}$ and the
global spline coefficients $\boldsymbol\varpi$, the problem decomposes into
$K$ independent local-position updates and one global spline-coefficient update, as illustrated in Fig~\ref{fig:lifted_penalty_framework}. For fixed $\boldsymbol\varpi^{(i)}$, the local subproblems are
\begin{equation}
\mathrm P_{z_k}:\quad
\mathbf z_k^{*}=\underset{\mathbf z_k\in\mathbb R^2}{\arg\min}\;
J_k(\mathbf z_k)+\frac{\eta}{2}
\mathcal{Z}_{k}(\mathbf{z}_{k},\boldsymbol{\varpi}),
\label{eq:local_subproblem}
\end{equation}
Once the $\{\mathbf z_k^{(i+1)}\}_{k=1}^{K}=\{\mathbf z_k^{*}\}_{k=1}^{K}$ have been updated, they are fixed for the global spline-coefficient update:
\begin{equation}
\begin{aligned}
\mathrm P_{\boldsymbol\varpi}:~
\boldsymbol\varpi^{*}=\underset{\boldsymbol\varpi}{\arg\min}\Bigg[
\mathcal{R}(\boldsymbol\varpi)
+\frac{\eta}{2}\sum_{k=1}^{K}
\mathcal{Z}_{k}(\mathbf{z}_{k},\boldsymbol{\varpi})\Bigg].
\end{aligned}
\label{eq:global_subproblem}
\end{equation}


\section{Alternating Solver for the Lifted Motion-Native Penalty Problem}

Building on the lifted objective and block subproblems in
\eqref{eq:lifted cost function}--\eqref{eq:global_subproblem}, this section
specifies their numerical solution. We first present the local and global updates of one alternating
cycle, then exploit their parallel and B-spline-induced structure for
implementation and complexity analysis, and finally state the fixed-stage descent property.

\subsection{Alternating Local and Global Updates}

At alternating cycle $i$, the $K$ local subproblems are solved in parallel with $\varpi^{(i)}$ fixed, using coarse search followed by safeguarded LM refinement.

For the coarse stage, let
$\Omega\subset\mathbb R^2$ be a finite set of sampled offsets around the
current spline position. The corresponding candidate set is
\begin{equation}
	\mathcal G_k
	\triangleq
	\left\{
	\mathbf S_k\boldsymbol\varpi^{(i)}
	+\Delta\mathbf z:
	\Delta\mathbf z\in\Omega
	\right\}
	\cup
	\left\{
	\mathbf z_k^{(i)}
	\right\},
\end{equation}
whose first part samples nearby positions and second part retains the incumbent local variable. The coarse initializer is then selected by evaluating the actual local
objective over this finite set to
provide a reliable starting point for the subsequent safeguarded LM
refinement:
\begin{equation}
	\mathbf z_k^{(i,0)}
	=
	\underset{\mathbf z\in\mathcal G_k}{\arg\min}
	\left[
	J_k(\mathbf z)
	+
	\frac{\eta}{2}
	\left\|
	\mathbf z
	-
	\mathbf S_k\boldsymbol\varpi^{(i)}
	\right\|_2^2
	\right].
	\label{eq:coarse_candidate_selection}
\end{equation}

Starting from $\mathbf z_k^{(i,0)}$, define the profiled residual and its
position Jacobian as $\mathbf r_k(\mathbf z)
\triangleq
\mathbf y_k-
\alpha_k(\mathbf z)\boldsymbol\psi_k(\mathbf z)$ and $\mathbf J_{r,k}(\mathbf z)
\triangleq
\nabla_{\mathbf z}\mathbf r_k(\mathbf z)$. The gradient and Gauss--Newton curvature of the local objective are
then modeled by
\begin{equation}
	\begin{aligned}
		\mathbf g_k
		&=
		\frac{2}{\sigma^2}
		\operatorname{Re}
		\left\{
		\mathbf J_{r,k}^{\mathsf H}\mathbf r_k
		\right\}
		+
		\eta
		\left(
		\mathbf z-\mathbf S_k\boldsymbol\varpi^{(i)}
		\right),                                                     \\[0.4em]
		\mathbf H_k
		&=
		\frac{2}{\sigma^2}
		\operatorname{Re}
		\left\{
		\mathbf J_{r,k}^{\mathsf H}\mathbf J_{r,k}
		\right\}
		+
		\eta\mathbf I_2 .
	\end{aligned}
	\label{eq:local_gn_model}
\end{equation}
All position-dependent quantities in
\eqref{eq:local_gn_model} are evaluated at the current local iterate. Given
a damping parameter $\mu\geq0$, the LM candidate and the safeguarded update
are
\begin{equation}
	\begin{aligned}
		\widetilde{\mathbf z}_k
		&=
		\mathbf z-
		\left(
		\mathbf H_k+\mu\mathbf I_2
		\right)^{-1}\mathbf g_k,                                    \\[0.4em]
		\mathbf z_k^{+}
		&\in
		\underset{
			\mathbf u\in\{\mathbf z,\widetilde{\mathbf z}_k\}
		}{\arg\min}\;
		\left[
		J_k(\mathbf u)
		+
		\frac{\eta}{2}
		\mathcal{Z}_{k}(\mathbf{u},\boldsymbol{\varpi})
		\right].
	\end{aligned}
	\label{eq:safeguarded_local_update}
\end{equation}
The second line evaluates the actual local objective and retains the current
iterate whenever the LM candidate fails to decrease it. After local
refinement terminates, we set
$\mathbf z_k^{(i+1)}=\mathbf z_k^{+}$. The same procedure is carried out
independently for all $k$, enabling parallel execution.

After the parallel local updates, fixing
$\{\mathbf z_k^{(i+1)}\}_{k=1}^{K}$ yields a quadratic global
subproblem in $\boldsymbol\varpi$. Setting the gradient of \eqref{eq:global_subproblem} with respect to $\boldsymbol\varpi$ to zero gives
\begin{equation}
	\mathbf C_{\boldsymbol\varpi}
	\boldsymbol\varpi^{(i+1)}
	=
	\boldsymbol\Gamma_{\boldsymbol\varpi}^{(i+1)},
	\label{eq:global_linear_system}
\end{equation}
where the corresponding global normal matrix $\mathbf C_{\boldsymbol\varpi}\in\mathbb R^{2L\times2L}$ and right-hand side vector $\boldsymbol\Gamma_{\boldsymbol\varpi}^{(i+1)}\in\mathbb R^{2L}$ are specified as
\begin{equation}
	\begin{aligned}
		\mathbf C_{\boldsymbol\varpi}
		&\triangleq
		\mathbf R
		+
		\eta\sum_{k=1}^{K}
		\mathbf S_k^{\mathsf T}\mathbf S_k,
		\label{eq:global_normal_matrix}\\		
		\boldsymbol\Gamma_{\boldsymbol\varpi}^{(i+1)}
		&\triangleq
		\eta\sum_{k=1}^{K}
		\mathbf S_k^{\mathsf T}\mathbf z_k^{(i+1)}\in\mathbb R^{2L}.
	\end{aligned}
\end{equation}
When $\mathbf C_{\boldsymbol\varpi}\succ\mathbf0$,
\eqref{eq:global_linear_system} admits a unique global minimizer. 
Algorithm~\ref{alg:lifted_map} summarizes the complete alternating solver. We
next examine how the B-spline structure determines the fast implementation and reduces
computational cost.

\begin{algorithm}[t]
	\caption{Alternating Solver for the Lifted Penalty Surrogate}
	\label{alg:lifted_map}
	\begin{algorithmic}[1]
		\Require $\{\mathbf y_k,t_{n_k},\gamma_k\}_{k=1}^{K}$; fixed $\mathbf W$;
		$\{\mathbf S_k\}$, $\mathbf R$, $\{\eta_s\}_{s=1}^{S}$; solver tolerances.
		\Ensure $\widehat{\boldsymbol\varpi}$,
		$\widehat{\mathbf p}(t)=
		\mathbf S(t)\widehat{\boldsymbol\varpi}$, and the reconstructed continuous
		channel $\widehat{\mathbf h}(t)$.
		
		\State Initialize $\boldsymbol\varpi$ and $\{\mathbf z_k\}$.
		\For{$s=1,\ldots,S$}
		\State Set $\eta\gets\eta_s$.
		\For{$i=0,\ldots,I_{\max,s}-1$}
		\For{$k=1,\ldots,K$ in parallel}
		\State Select the coarse candidate using
		\eqref{eq:coarse_candidate_selection}.
		\State Refine using
		\eqref{eq:local_gn_model} and apply the safeguard
		\eqref{eq:safeguarded_local_update}.
		\State Set the accepted result as $\mathbf z_k^{(i+1)}$.
		\EndFor
		\State Form
		$\mathbf C_{\boldsymbol\varpi}$ and
		$\boldsymbol\Gamma_{\boldsymbol\varpi}^{(i+1)}$, and solve
		\eqref{eq:global_linear_system}.
		\If{the fixed-stage stopping tests pass}
		\State \textbf{break}.
		\EndIf
		\EndFor
		\EndFor
		\State Reconstruct
		$\widehat{\mathbf p}(t)=
		\mathbf S(t)\widehat{\boldsymbol\varpi}$.
		\State Recover $\{\widehat{\alpha}_k\}_{k=1}^{K}$ and reconstruct $\widehat{\mathbf h}(t)$.
		\State \Return
		$\widehat{\mathbf h}(t)$.
	\end{algorithmic}
\end{algorithm}

\subsection{Structured Fast Implementation}

We next exploit the local support of the B-spline basis and the shared
Kronecker structure to characterize
the resulting global sparsity. The required properties are established
in the following two lemmas, with proofs provided in
Appendix~\ref{app:bspline_structural_proofs}.

\begin{lemma}[Local support of the B-spline basis]
	\label{lem:bspline_local_support}
	For the degree-$q$ B-spline basis associated with the knot
	$\{\tau_i\}$, its support satisfies $\operatorname{supp}\!\left(B_{\ell,q}\right)
	\subseteq
	[\tau_\ell,\tau_{\ell+q+1}],
	\ell=0,\ldots,L-1.$
	Hence, at most $q+1$ consecutive basis functions are nonzero.
\end{lemma}

\begin{lemma}[Banded sparsity of the global normal matrix]
	\label{lem:global_banded_sparsity}
	The matrix $\mathbf C_{\boldsymbol\varpi}$ is block banded with block
	half-bandwidth
	$b_{\mathbf C}\triangleq\max\{q,2\}$
	for $\ell,r=0,\ldots,L-1$, independently of $L$.
\end{lemma}

As Lemmas~\ref{lem:bspline_local_support} and
\ref{lem:global_banded_sparsity} reveal the banded structure of
$\mathbf C_{\boldsymbol\varpi}$, we further exploit the common
Kronecker structure already embedded in $\mathbf S_k$ and
$\mathbf D_x$, which reduces the $2L$-dimensional global update to two identical $L$-dimensional systems. Specifically, we have $\mathbf C_{\boldsymbol\varpi}=\mathbf C_s(\eta)\otimes\mathbf I_2$, where
\begin{equation}
	\mathbf C_s(\eta)
	\triangleq
	\eta\sum_{k=1}^{K}
	\mathbf b(t_{n_k})\mathbf b^{\mathsf T}(t_{n_k})
	+
	\lambda_1\boldsymbol\Delta_1^{\mathsf T}\boldsymbol\Delta_1
	+
	\lambda_2\boldsymbol\Delta_2^{\mathsf T}\boldsymbol\Delta_2.
	\label{eq:scalar_global_matrix}
\end{equation}
Since each $\mathbf b(t_{n_k})$ has at most $q+1$ active entries
and the difference operators are naturally banded,
$\mathbf C_s(\eta)$ also has a fixed bandwidth independent of $L$.
Therefore, the two trajectory coordinates share the same banded
system matrix, so the global update requires no explicit matrix
inverse. 

Rather than explicitly forming \(\mathbf C_{\varpi}^{-1}\),
we factorize the scalar banded matrix as
$\mathbf C_s(\eta)=\mathbf L_s\mathbf L_s^{\mathsf T}$,
where $\mathbf L_s$ is the triangular banded Cholesky factor, and each global update is
computed by two banded triangular solves without explicitly forming $\mathbf C_{\boldsymbol\varpi}^{-1}$:
\begin{equation}
	\begin{aligned}
		(\mathbf L_s\otimes\mathbf I_2)
		\mathbf v^{(i+1)}
		&=
		\boldsymbol\Gamma_{\boldsymbol\varpi}^{(i+1)},\\
		(\mathbf L_s^{\mathsf T}\otimes\mathbf I_2)
		\boldsymbol\varpi^{(i+1)}
		&=
		\mathbf v^{(i+1)},
	\end{aligned}
	\label{eq:factorized_global_solve}
\end{equation}
where $\mathbf v^{(i+1)}$ is only the intermediate result of the forward substitution. Moreover, $\mathbf C_s(\eta)$ remains fixed within each continuation
stage, so its $\mathcal O(Lb_{\mathbf C}^2)$ banded Cholesky
factorization is performed only once and reused across alternating cycles.
For fixed spline degree, $b_{\mathbf C}$ is constant, making both the
factorization and repeated solves linear in $L$.

\subsection{Algorithmic Efficiency and Fixed-Stage Descent}

We now quantify the ideal parallel local-block cost, the banded global-block cost, and the fixed-stage descent property.

\begin{table*}[t]
	\caption{Complexity before and after exploiting the local--global
		structure.}
	\label{tab:complexity_comparison}
	\centering
	\footnotesize
	\renewcommand{\arraystretch}{1.22}
	\setlength{\tabcolsep}{5pt}
	
	\begin{tabularx}{0.97\textwidth}{
			@{}
			p{0.22\textwidth}
			X
			X
			@{}
		}
		\toprule
		\textbf{Computational component}
		&
		\textbf{Serial and dense reference}
		&
		\textbf{Parallel and banded implementation}
		\\
		\midrule
		
		Local block per cycle
		&
		$\mathcal O(KMN_{\rm RF})$
		&
		$\mathcal O(MN_{\rm RF})$
		\\
		
		Global setup and one solve
		&
		$\mathcal O(KL^2+L^3)$
		&
		$\mathcal O\!\left(
		K(q+1)^2+Lb_{\mathbf C}^2
		\right)$
		\\
		
		Repeated global update
		&
		$\mathcal O(KL+L^2)$
		&
		$\mathcal O\!\left(
		K(q+1)+Lb_{\mathbf C}
		\right)$
		\\
		
		Complete alternating cycle
		&
		$\mathcal O\!\left(
		KMN_{\rm RF}+KL+L^2
		\right)$
		&
		$\mathcal O\!\left(
		MN_{\rm RF}+K(q+1)+Lb_{\mathbf C}
		\right)$
		\\
		\bottomrule
		
	\end{tabularx}
	\vspace{1mm}
	\parbox{\columnwidth}{
		\footnotesize
		\textit{Note:} For the local block, the parallel entry denotes ideal execution latency rather than total arithmetic work.}
\end{table*}

\subsubsection{Local-Block Complexity}

With $\boldsymbol\varpi^{(i)}$ fixed, the $K$ local subproblems are
independent. Each refinement is dominated by evaluating the dense-combined
near-field response and its position derivatives, requiring
$\mathcal O(MN_{\rm RF})$ time. Serial execution therefore scales as
$\mathcal O(KMN_{\rm RF})$, whereas concurrent execution reduces the ideal
local-block time to
\begin{equation}
	T_{\rm loc}^{(\mathrm{cycle})}
	=
	\mathcal O(MN_{\rm RF}).
	\label{eq:local_cycle_complexity}
\end{equation}
Thus, parallel execution removes the factor $K$ from the ideal local-block latency.

\subsubsection{Global-Block and Complete-Cycle Complexity}

Without exploiting the B-spline locality and Kronecker structure,
dense assembly and factorization of the global system require
$\mathcal O(KL^2+L^3)$ operations. In the structured implementation, computing the banded Cholesky factorization\cite{golub2013matrix} requires $\mathcal O\!\left(K(q+1)^2+Lb_{\mathbf C}^2
\right)$. 

Within each continuation stage, $\mathbf C_s(\eta)$ remains fixed and
is therefore factorized only once. Each subsequent global update requires
$\mathcal O(K(q+1))$ operations for right-hand-side assembly and
$\mathcal O(Lb_{\mathbf C})$ operations for the banded forward--backward
substitutions. The banded representation also reduces the global-system
storage from $\mathcal O(L^2)$ to $\mathcal O(Lb_{\mathbf C})$.
Combining this global update with the ideal parallel local block gives
\begin{equation}
	T^{(\mathrm{cycle})}
	=
	\mathcal O\!\left(
	MN_{\rm RF}
	+
	K(q+1)
	+
	Lb_{\mathbf C}
	\right).
	\label{eq:cycle_complexity}
\end{equation}
Hence, parallelization removes the factor $K$ from the ideal local-block
latency, while the fixed-bandwidth triangular solves reduce the repeated
global update from quadratic to linear scaling in $L$, as summarized in
Table~\ref{tab:complexity_comparison}.

Importantly, the efficiency gains summarized above do not modify the
underlying block updates. Parallelization only changes the execution order of
the mutually independent local subproblems, while the banded implementation
solves the same global normal equations as dense linear algebra. The
structured implementation therefore preserves the descent property of the
original alternating scheme.

For fixed $\eta$, the safeguarded local updates are nonincreasing and the global step exactly minimizes the quadratic
$\boldsymbol\varpi$-subproblem. Hence, each alternating cycle satisfies
\begin{equation}
	\begin{aligned}
		J_\eta\!\left(
		\boldsymbol\varpi^{(i+1)},\{\mathbf z_k^{(i+1)}\}
		\right)
		&\le
		J_\eta\!\left(
		\boldsymbol\varpi^{(i)},\{\mathbf z_k^{(i+1)}\}
		\right)\\
		&\le
		J_\eta\!\left(
		\boldsymbol\varpi^{(i)},\{\mathbf z_k^{(i)}
		\}\right).
	\end{aligned}
	\label{eq:cycle_descent}
\end{equation}
Therefore, the structured parallel implementation preserves monotonic descent. Since \(J_\eta\) is bounded below, its objective-value sequence converges within each continuation stage. This provides a clear fixed-stage stability guarantee for the proposed alternating solver. However, objective-value convergence does not by itself establish convergence of the iterates to a stationary point or a global optimum of the underlying nonconvex problem. Establishing such stronger guarantees requires substantially deeper theoretical analysis and constitutes an important direction for future work.
\section{Simulation and Validation}
\label{sec:sim}

\subsection{Simulation Setup and Comparison Overview}

To ensure a fair comparison, all methods are evaluated under
the same physical geometry, observation model, and evaluation protocol, so
that the resulting performance differences reflect estimator and solver design.

\subsubsection{Simulation Configuration}

Unless otherwise stated, all experiments use the nominal settings in Table~\ref{tab:sim-params}, with following study-specific variations reported in the
corresponding subsections. Throughout the simulations, the reported SNR denotes the nominal pre-combining per-antenna pilot SNR. The BS is located at $[0,0]^{\mathsf T}$, and the
centered ULA is aligned with the global $x$-axis. The aperture is $D=(M-1)\lambda/2=382.5$~mm, yielding the Rayleigh distance
\begin{equation}
	R_R=\frac{2D^2}{\lambda}\approx97.5~\mathrm{m}.
	\label{eq:rayleigh_distance}
\end{equation}
The principal experiments satisfy $r_{\max}<R_R$, except for the distance sweep, which crosses $R_R$ to evaluate tracking performance around the conventional
near--far-field boundary.
\begin{table}[t]
	\centering
	\caption{Main simulation and solver settings.}

	\renewcommand{\arraystretch}{1.10}
	\small
	\setlength{\tabcolsep}{3pt}
	\begin{tabularx}{\columnwidth}{
			@{}
			>{\raggedright\arraybackslash}p{0.35\columnwidth}
			>{\centering\arraybackslash}p{0.12\columnwidth}
			>{\raggedright\arraybackslash}X
			@{}
		}
		\toprule
		\textbf{Parameter}
		& \textbf{Symbol}
		& \textbf{Value} \\
		\midrule		
		Carrier frequency
		& $f_c$
		& $100$~GHz \\
		
		Number of antennas
		& $M$
		& $256$ \\
		
		Number of RF chains
		& $N_{\rm RF}$
		& $16$ to $256$(default) \\
		
		Element spacing
		& $d$
		& $1.5$~mm $(=\lambda/2)$ \\
		
		Observation duration
		& $T$
		& $20$~s \\
		
		Number of timeslots
		& $N$
		& $200$ \\
		
		Number of pilot slots
		& $K$
		& $10$ \\
		
		Pilot symbol
		& $\gamma_k$
		& $1+\mathrm j$ \\
		
		Default SNR
		& SNR
		& $20$dB \\
		
		Spline degree
		& $q$
		& $3$ \\
		
		Control points Number
		& $L$
		& $8$ \\
		
		Knot sequence
		& $\{\tau_\ell\}$
		& Uniform \\
		
		Regularization weights
		& $\lambda_1,\lambda_2$
		& $10^{-6},10^{-4}$ \\
		
		Continuation schedule
		& $\eta_s$
		& $10^{-2}$(default) to $20$ \\
		
		Coarse-search offset set
		& $\Omega$
		& $5\times 5$ grids over $[-10,10]^2$~m \\
		
		Global initialization
		& $\boldsymbol\varpi^{(0)}$
		& Interpolated straight line \\
		
		Maximum cycles
		& $I_{\max,s}$
		& $16$ \\
		
		\bottomrule
	\end{tabularx}
	\label{tab:sim-params}
\end{table}

\subsubsection{Compared Baselines}

To compare with conventional estimate-then-track pipelines, we consider three
snapshot estimators and three temporal-processing strategies.

The snapshot-estimation stage includes:
\begin{itemize}
	\item \textbf{On-grid OMP} \cite{cui2022channel}, which performs
	polar-domain sparse recovery over a predefined dictionary;
	
	\item \textbf{Off-grid P-SIGW} \cite{cui2022channel}, which refines the
	spatial parameters in the continuous domain to mitigate grid mismatch;
	and
	
	\item \textbf{Joint ML} \cite{wang2024near}, which directly estimates the
	continuous position at each pilot instant under the adopted observation
	model.
\end{itemize}

The temporal-processing stage includes:
\begin{itemize}
	\item \textbf{Raw output}, which retains the independently estimated
	pilot positions;
	
	\item \textbf{LS B-spline smoothing} \cite{eilers1996flexible}, which fits
	a continuous trajectory to the position snapshots; and
	
	\item \textbf{RTS smoothing} \cite{zhang2026near}, which serves as a
	state-space smoothing reference.
\end{itemize}

The resulting baselines are OMP+Raw, P-SIGW+Raw, Joint-ML+Raw, OMP+RTS, OMP+LS, P-SIGW+LS, and Joint-ML+LS. These pipelines estimate discrete snapshots before imposing temporal consistency, whereas the proposed method directly couples all pilot observations. 

To ensure a fair comparison, all temporally refined methods receive the same ground-truth start and end positions as common boundary side information, so the comparison focuses on interior trajectory reconstruction rather than boundary initialization. For the proposed method, these endpoints are imposed through the quadratic anchor
\begin{equation}
	J_{\rm anc}(\boldsymbol\varpi)
	=
	\frac{\kappa_{\rm anc}}{2}
	\sum_{r\in\{\rm s,e\}}
	\left\|
	\mathbf S(t_r)\boldsymbol\varpi-\mathbf p_r^\star
	\right\|_2^2,
	\label{eq:simulation_endpoint_anchor}
\end{equation}
where $t_{\rm s}=t_1$, $t_{\rm e}=t_N$, and
$\kappa_{\rm anc}=10^6$. In the simulations, $J_{\rm anc}(\boldsymbol\varpi)$ is added to the global spline subproblem in equation \eqref{eq:global_subproblem}, equivalently, it contributes the corresponding fixed quadratic terms to $\mathbf C_{\varpi}$ and $\boldmath \Gamma_{\varpi}$, without altering the local updates or the banded structure of the solver.
LS baselines impose the same endpoints as equality constraints, whereas RTS uses them as high-confidence boundary updates.

\subsubsection{Common Comparison and Evaluation Protocol}

In the following experiments, $\mathbf W$ is constructed from the
$N_{\rm RF}$ rows of a fixed QR-orthogonalized complex-Gaussian matrix,
yielding nested dense semiunitary combiners shared by all methods.
Unless otherwise stated, the reported results are executed over $N_{\rm MC}=60$ paired Monte Carlo trials, which reduces
the influence of occasional large-error realizations.

\subsection{Channel Tracking Accuracy}

\begin{figure}[t]
	\centering
	\includegraphics[width=3in,keepaspectratio]
	{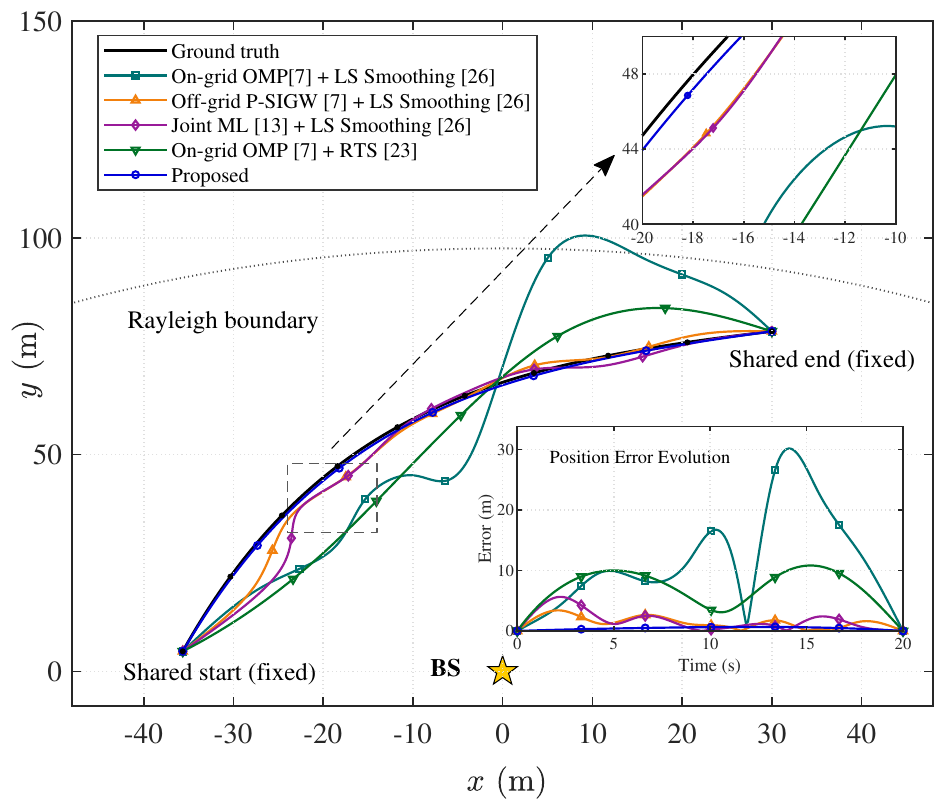}
	\caption{Trajectory reconstruction and pointwise position error for one
		representative realization.}
	\label{Fig_Sim01}
\end{figure}
\begin{figure}[t]
	\centering
	\includegraphics[width=3.3in,keepaspectratio]
	{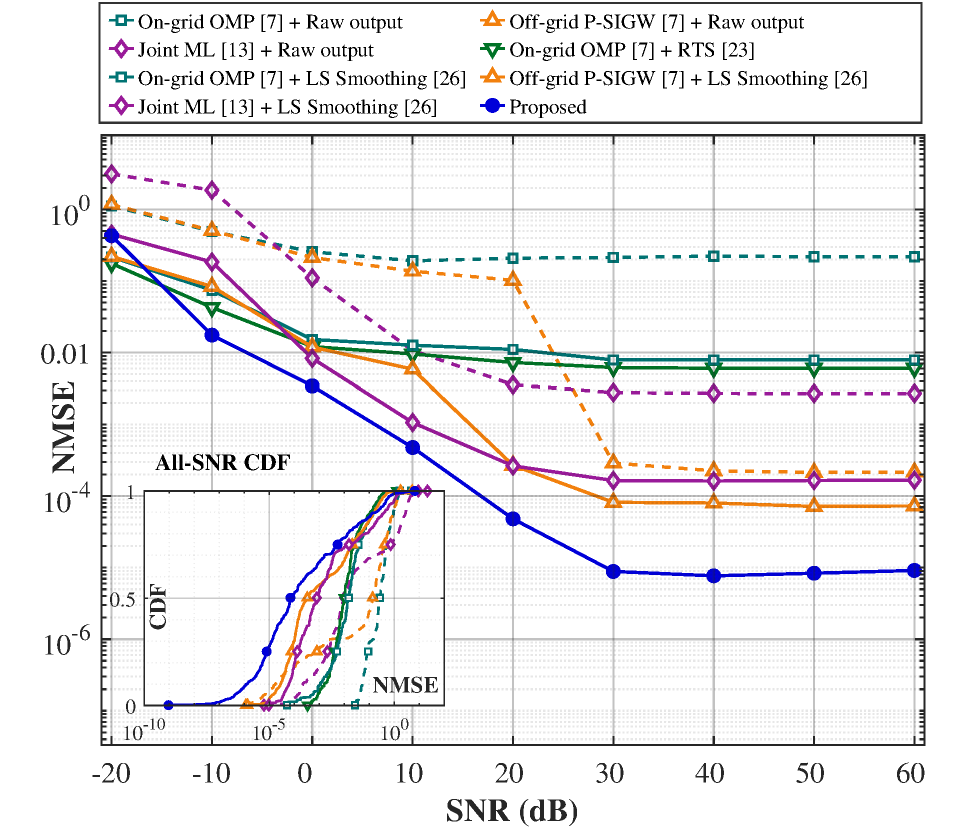}
	\caption{NMSE versus SNR with the pooled empirical CDF over the tested
		SNR range.}
	\label{Fig_Sim03}
\end{figure}
\begin{figure}[t]
	\centering
	\includegraphics[width=3in,keepaspectratio]
	{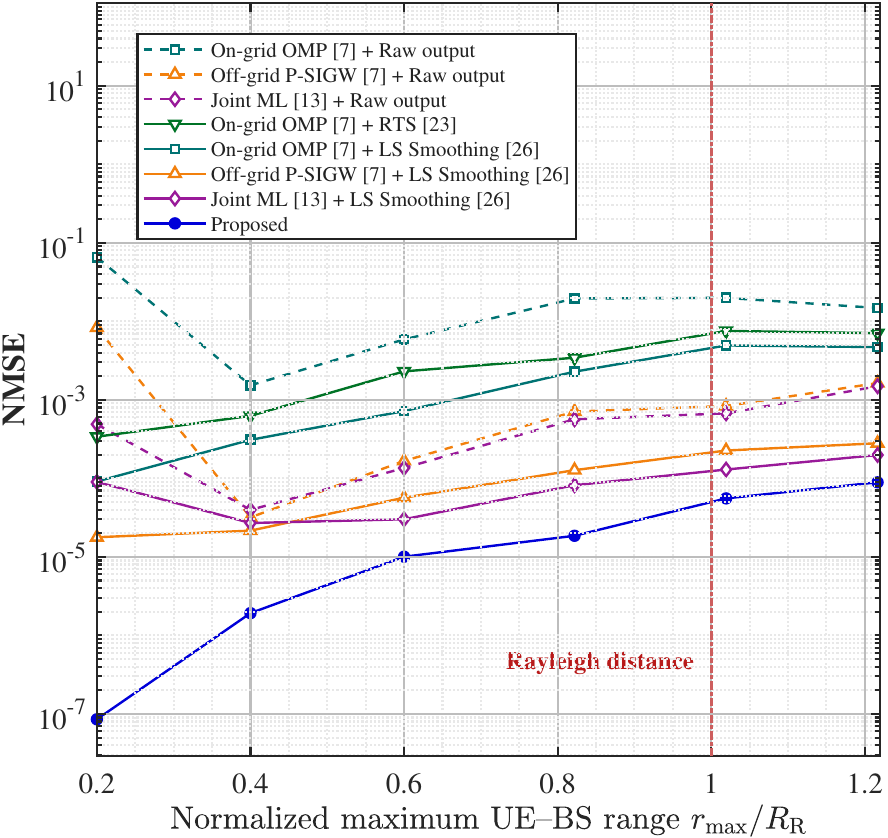}
	\caption{NMSE versus normalized maximum UE–BS range.}
	\label{Fig_DistanceNMSE}
\end{figure}

We first evaluate the proposed framework through one representative trajectory
realization and two statistical NMSE studies. Specifically, Fig.~\ref{Fig_Sim01} presents a representative trajectory-level comparison and shows how different methods reconstruct the
channel evolution between sparse pilot instants.
Fig.~\ref{Fig_Sim03} evaluates the NMSE over a broad SNR range, while
Fig.~\ref{Fig_DistanceNMSE} examines the NMSE at different propagation
distances. Together, these experiments characterize both representative and statistical tracking performance.

As shown in Fig.~\ref{Fig_Sim01}, all temporally refined methods share the
same endpoints under the common comparison protocol. The proposed method
more accurately reconstructs the interior trajectory and maintains a
consistently low position error, whereas the snapshot-first pipelines exhibit
larger local deviations.

Fig.~\ref{Fig_Sim03} further evaluates the channel-tracking performance over a broad
SNR range. As the SNR increases, the proposed method continues to reduce the NMSE
and achieves a clear advantage over the snapshot-first baselines. This
trend is consistent with the information perspective in Sec.~III-B. When the pilot observations become sufficiently informative, directly
aggregating their geometric constraints through the shared trajectory
avoids committing prematurely to independently estimated snapshots. In contrast, errors
introduced during pilot-local estimation and subsequent temporal
processing remain in the snapshot-first pipelines. The pooled empirical
CDF over the tested SNR range further shows that the proposed method is
consistently concentrated toward lower NMSE values, rather than obtaining
its gain only at a few isolated SNR points.

Fig.~\ref{Fig_DistanceNMSE} reports the NMSE from $0.2R_R$ to
$1.2R_R$ at $20$~dB. The proposed method maintains the lowest NMSE
among the temporally refined methods throughout the tested range, while
its error increases gradually with distance. This behavior follows the
near-field observation model in Sec.~II: increasing range progressively
weakens the spherical-wave curvature that encodes position, thereby
reducing geometric sensitivity. Accordingly, no abrupt performance change
is observed at the conventional Rayleigh boundary $R_R$.

\subsection{Performance under Representative Operating Scenarios}

We next examine the proposed framework under three representative
variations: increasing motion intensity, changing spatial and temporal
observation support, and propagation-model mismatch. These experiments
respectively probe the effects of trajectory-model bias, available
geometric information, and unmodeled coherent residuals induced by an NLoS path.

\begin{figure}[t]
	\centering
	\includegraphics[width=3in,keepaspectratio]
	{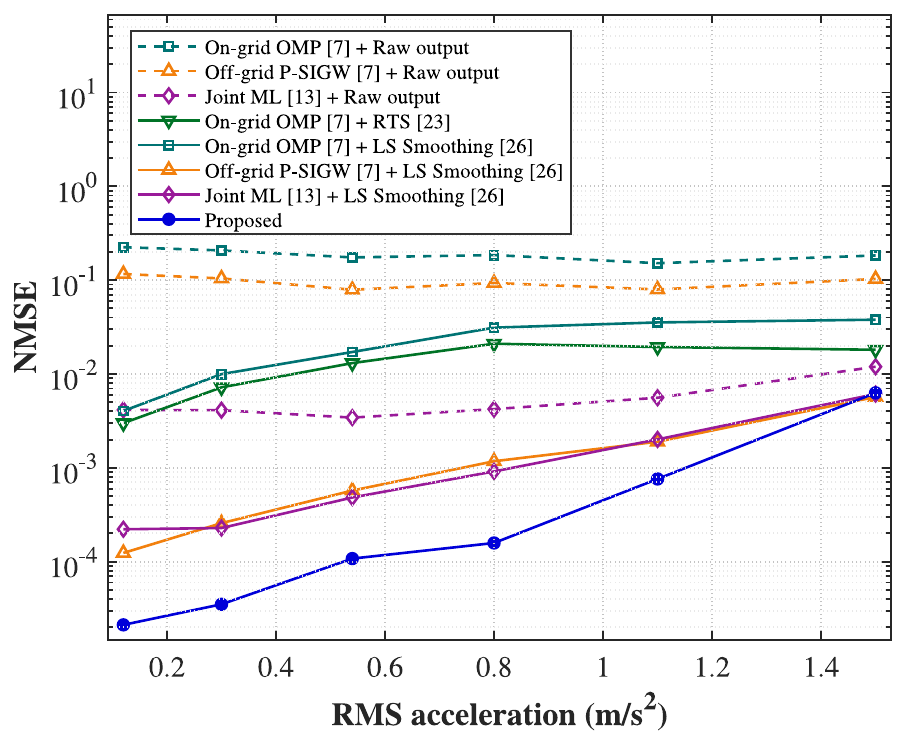}
	\caption{NMSE versus RMS acceleration.}
	\label{Fig_Sim05}
\end{figure}

\textbf{1) Smooth accelerated motion:}
We first vary the trajectory complexity through the RMS acceleration
\begin{equation}
	a_{\rm RMS}
	=
	\left[
	\frac{1}{t_{\rm e}-t_{\rm s}}
	\int_{t_{\rm s}}^{t_{\rm e}}
	\left\|
	\frac{\mathrm d^2\mathbf p(t)}{\mathrm dt^2}
	\right\|_2^2
	\mathrm dt
	\right]^{1/2}.
	\label{eq:rms_acceleration}
\end{equation}
As shown in Fig.~\ref{Fig_Sim05}, the proposed method achieves the lowest NMSE for $a_{\rm RMS}\leq1.4~\mathrm{m/s^2}$. However, when the motion
becomes more complex, the fixed smoothness prior increasingly suppresses genuine trajectory variations, leading to a larger regularization bias. This limitation is also partly associated with the adopted open-uniform
knot sequence, which allocates the spline resolution uniformly over time
regardless of the local motion intensity. A natural extension is therefore to adopt nonuniform or adaptive knot placement, assigning denser spline support to intervals with stronger motion variations in future work.

\begin{figure}[t]
	\centering
	\includegraphics[width=3in,keepaspectratio]
	{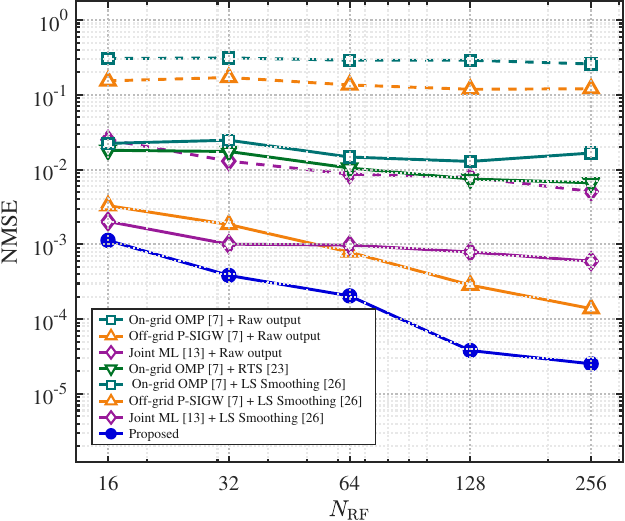}
	\caption{NMSE versus combiner output dimension.}
	\label{Fig_Sim06}
\end{figure}

\begin{figure}[t]
	\centering
	\includegraphics[width=3in,keepaspectratio]
	{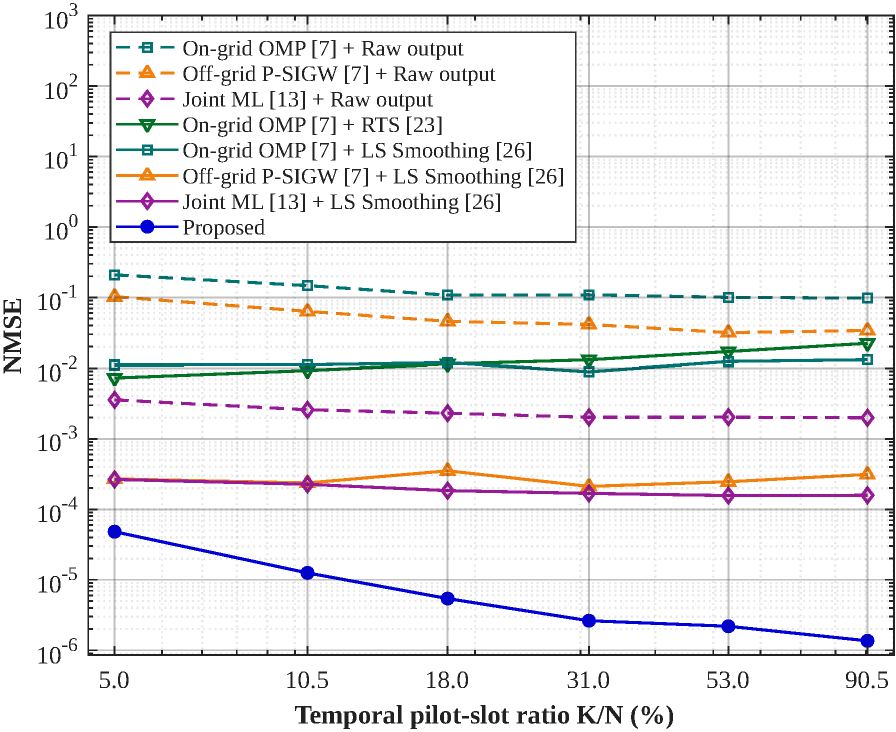}
	\caption{NMSE versus temporal pilot-slot ratio.}
	\label{Fig_Sim07}
\end{figure}

\textbf{2) Spatial and temporal observation support:}
Figs.~\ref{Fig_Sim06} and~\ref{Fig_Sim07} examine how the proposed method
benefits from increased spatial and temporal observation support,
respectively.

In Fig.~\ref{Fig_Sim06}, increasing $N_{\rm RF}$ consistently
improves the proposed method. Since the nested semiunitary
combiners preserve the physical aperture, this gain results from
retaining more post-combining response dimensions. As characterized
by equations \eqref{eq:gain_orthogonal_projector},\eqref{eq:single_pilot_spline_information},
only the geometry-sensitive response variations orthogonal to the
nuisance direction of $\alpha_k$ contribute efficient trajectory
information. Increasing $N_{\rm RF}$ preserves more such components,
thereby strengthening the pilot-local information and its mapped
constraint on the coefficient vector.

Fig.~\ref{Fig_Sim07} further varies the temporal pilot support by
increasing $K$. The proposed method maintains the lowest NMSE over the
tested range and generally improves as the pilot density increases.
Unlike Fig.~\ref{Fig_Sim06}, which strengthens the information supplied
by each individual pilot, increasing $K$ introduces more pilot-local
contributions into equation \eqref{eq:total_coefficient_information}
and provides denser temporal sampling of the shared spline. According to
Proposition~\ref{prop:weighted_sampling_identifiability}, these additional
mapped contributions improve the coverage and conditioning of the shared
coefficient space, thereby reducing the trajectory estimation error.

\textbf{3) Unmodeled NLoS-path mismatch:}
We then examine propagation-model mismatch by adding one unmodeled NLoS reflected path to the observations while all methods retain the LoS-based likelihood. The NLoS-to-LoS power ratio is varied from the LoS-only case to $-5$~dB.

\begin{figure}[t]
	\centering
	\includegraphics[width=3in,keepaspectratio]
	{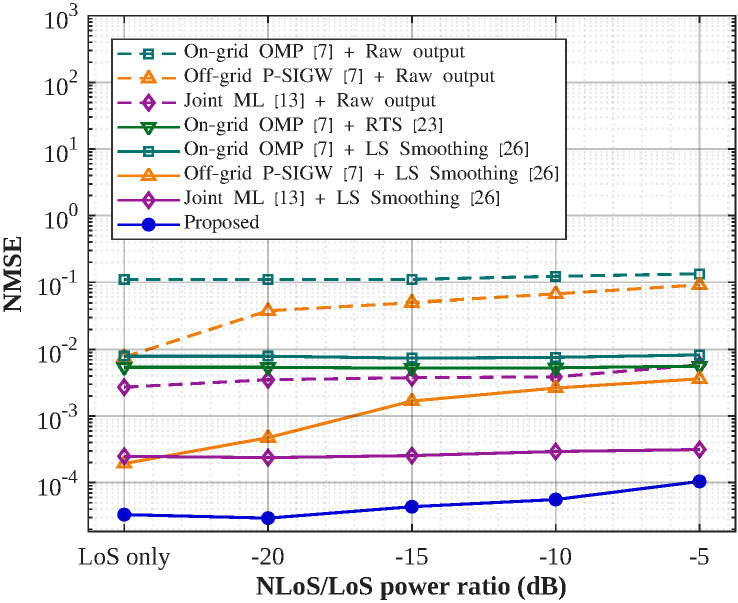}
	\caption{NMSE versus NLoS-to-LoS power ratio with one unmodeled reflected path.}
	\label{Fig_Sim09}
\end{figure}

As shown in Fig.~\ref{Fig_Sim09}, the proposed method maintains the
lowest NMSE over all tested mismatch levels. For weak reflections, the
shared coefficient vector prevents individual pilot positions from
independently drifting to absorb the unmodeled component, thereby
suppressing mismatch-induced local deviations. As the reflected power
increases, however, its residual component after nuisance projection
increasingly overlaps the geometry-sensitive directions characterized by
$\mathbf P_{\psi,k}^{\perp}\mathbf Q_{\boldsymbol\varpi,k}$. When such
components remain coherent across pilot instants, they accumulate in the shared coefficient space and gradually bias \(\boldsymbol\varpi\).

The proposed method nevertheless retains the lowest NMSE throughout the tested dominant-LoS mismatch range, with gradual degradation as the reflected component strengthens.

\subsection{Algorithmic Properties and Efficiency}

We finally examine the numerical and computational behavior of the proposed solver from four aspects, including regularization sensitivity, optimization behavior, structural acceleration of the alternating updates, and the resulting end-to-end accuracy--runtime trade-off.

\begin{figure}[t]
	\centering
	\includegraphics[width=3in,keepaspectratio]
	{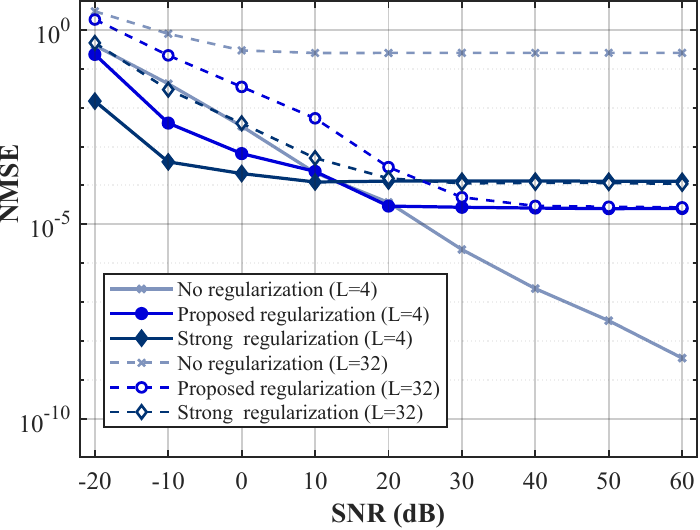}
	\caption{NMSE versus SNR under different regularization strengths and
		trajectory complexities.}
	\label{Fig_RegularizationSensitivity}
\end{figure}

\textbf{1) Regularization and trajectory complexity:}
We first examine how regularization interacts with trajectory complexity. Here, ground-truth and estimated trajectories use the same cubic B-spline family and the same $L$, so varying $L$ changes trajectory complexity without introducing representation mismatch. Fig.~\ref{Fig_RegularizationSensitivity}
compares three regularization strengths for trajectories with $L=4$ and
$L=32$ over the tested SNR range. The no-, nominal-, and strong-regularization settings correspond to $(\lambda_1,\lambda_2)=\mathbf 0$, the nominal values in
Table~\ref{tab:sim-params}, and $10^3$ times the nominal values respectively.

For \(L=4\), the default \(K=10\) satisfies the necessary dimension condition \(K\ge L\)  in
Proposition~\ref{prop:weighted_sampling_identifiability}, and under the tested pilot geometry the low-dimensional coefficient space is sufficiently constrained by the observations. Accordingly,
the unregularized estimator continues to improve with SNR without an
appreciable error floor, whereas regularization eventually introduces a
high-SNR bias floor.

In contrast, $L=32>K$ violates the necessary information-dimension
condition, making the observation-level coefficient information rank
deficient regardless of SNR. The unregularized estimator therefore
exhibits severe performance degradation, while regularization supplies
curvature along weakly or unsupported coefficient directions and
substantially stabilizes the estimation. Excessive regularization,
however, introduces additional smoothing bias, as reflected by the higher
floor of the strong-regularization setting.

Fig.~\ref{Fig_RegularizationSensitivity} therefore confirms the distinction in Sec. III-B. Observation-supported directions improve with SNR, whereas regularization mainly stabilizes unsupported directions at the cost of high-SNR bias.

\textbf{2) Optimization landscape and basin selection:}
We next examine the sensitivity to initialization and the role of the proposed lifted local--global formulation in navigating the nonconvex landscape.

\begin{figure}[t]
	\centering
	\includegraphics[width=3in,keepaspectratio]
	{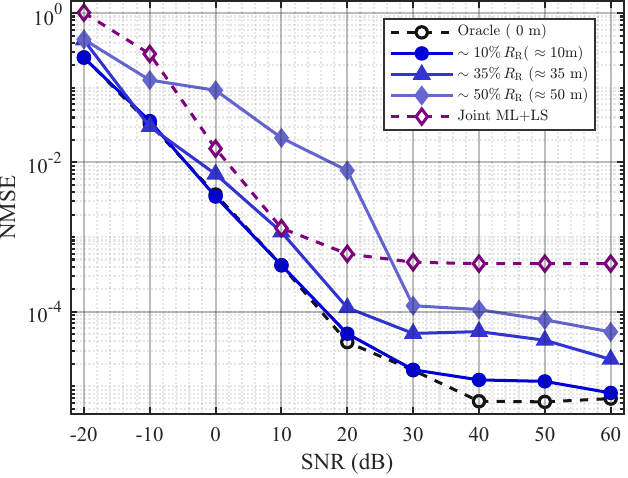}
	\caption{NMSE versus SNR under different initialization perturbations.}
	\label{Fig_InitializationNoise}
\end{figure}

\begin{figure}[t]
	\centering
	\includegraphics[width=\columnwidth,keepaspectratio]
	{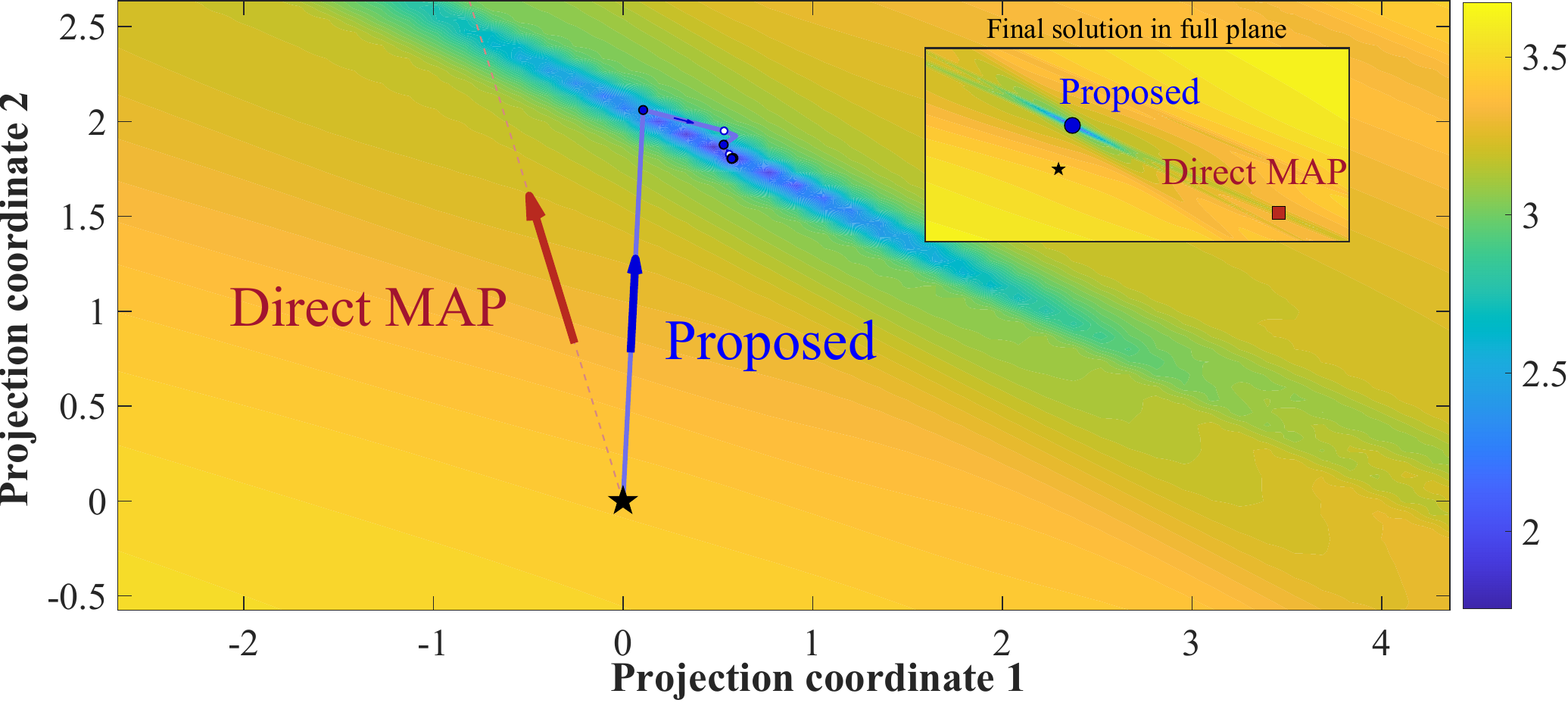}
	\caption{Solution-spanning slice of $\log_{10}J_{\rm dir}$ at
		$20$~dB, with the projected optimization paths of the lifted solver and
		safeguarded Direct MAP.}
	\label{Fig_AuxiliaryBasinSlice}
\end{figure}

Fig.~\ref{Fig_InitializationNoise} evaluates initialization offsets of
approximately $0.1R_R$, $0.35R_R$, and $0.5R_R$. As expected, larger
perturbations mainly degrade performance when the observations are weak.
With increasing SNR, however, even the $0.35R_R$ and $0.5R_R$
initializations recover substantially, while the $0.1R_R$ case approaches
the oracle initialization. This indicates that sufficiently informative
pilot observations can correct considerable initialization errors, whereas
weak observations combined with a distant initialization increase the
likelihood of convergence to unfavorable basins.

To further isolate the effect of variable lifting, we compare the proposed
solver with safeguarded Direct MAP, which minimizes
$J_{\rm dir}(\boldsymbol\varpi)$ directly from the same observations and
initialization but without the auxiliary variables or penalty continuation.
Fig.~\ref{Fig_AuxiliaryBasinSlice} projects their iterates onto a
two-dimensional solution-spanning subspace, with the background showing
$\log_{10}J_{\rm dir}$. Although both methods start from the same point, their optimization paths
rapidly diverge. Direct MAP follows the coupled objective directly and is
attracted to a different local basin, whereas the lifted solver follows the
local--global continuation toward the lower-objective valley and reaches a
more accurate terminal solution. This behavior illustrates how relaxing local--global consistency through the auxiliary variables at early continuation stages allows the pilot-local geometry to adjust before being progressively reconciled with the shared trajectory, thereby steering the solver toward a different basin of the nonconvex objective.

\textbf{3) Structured and parallel implementation efficiency:}

We next validate the two structural accelerations developed in Sec.~IV, namely, parallel local updates and the compact-banded global solve.

\begin{figure}[t]
	\centering
	\includegraphics[width=3.2in,keepaspectratio]
	{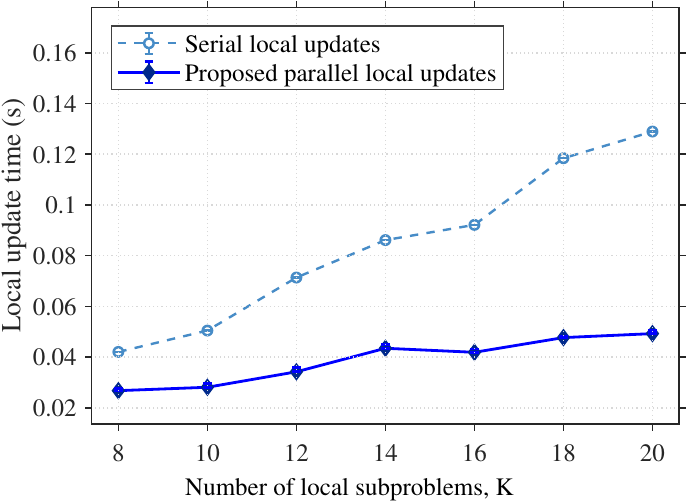}
	\caption{Local-block time under serial and parallel execution.}
	\label{Fig_LocalParallelTiming}
\end{figure}

Fig.~\ref{Fig_LocalParallelTiming} validates the local parallelization. As $K$ increases from $8$ to $20$, serial latency grows nearly linearly, whereas the parallel implementation increases slowly. At $K=20$, runtime decreases from $0.129$ s to $0.049$ s (2.6×).

\begin{figure}[t]
	\centering
	\includegraphics[width=3.1in,keepaspectratio]
	{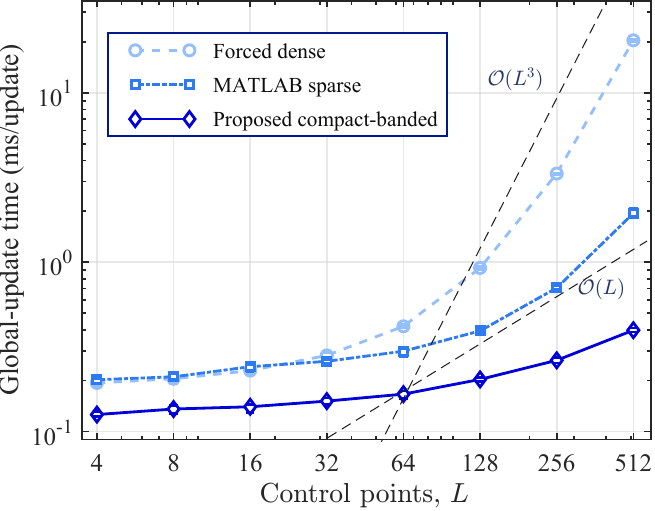}
	\caption{Time of each repeated global update versus control-point count.}
	\label{Fig_GlobalBackendTiming}
\end{figure}

Fig.~\ref{Fig_GlobalBackendTiming} measures the time of each repeated global update after the continuation-stage factorization is reused. Dense, generic-sparse, and compact-banded solvers are comparable for small $L$, but diverge as $L$ grows; the proposed backend follows the predicted near-linear scaling and becomes substantially cheaper at $L=512$.

Figs.~\ref{Fig_LocalParallelTiming} and
\ref{Fig_GlobalBackendTiming} experimentally confirm that parallelization suppresses the $K$-dependent
local latency, while fixed-bandwidth banded algebra prevents the global solve from growing quadratically or cubically with $L$. Their combination forms the computational basis for the end-to-end accuracy--runtime trade-off evaluated next.

\begin{figure}[t]
	\centering
	\includegraphics[width=3in]
	{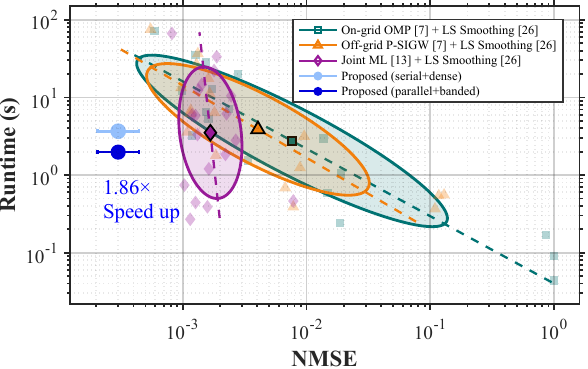}
	\caption{
		End-to-end runtime--NMSE operating regions at $10$~dB constructed from
		18 $(R_{sample},\beta)$ configuration-level mean points.
	}
	\label{Fig_RuntimeNMSEEnvelope}
\end{figure}

\textbf{4) End-to-end accuracy--runtime trade-off:}
The preceding experiments separately examine the numerical behavior and
the local--global structural efficiency of the proposed solver.
Fig.~\ref{Fig_RuntimeNMSEEnvelope} further compares the complete inference
pipelines in terms of end-to-end runtime and NMSE.
Since the SNR dependence has already been systematically characterized in
Fig.~\ref{Fig_Sim03}, we fix the SNR at the representative value of
$10$~dB here to isolate the accuracy--runtime variation induced by the
baseline search/codebook configurations.

For each snapshot-first baseline, following the polar-domain codebook
construction in~\cite{cui2022channel}, we vary the angular sampling rate
$R_{sample}\in\{1,2,4\}$ and the range-sampling parameter $\beta$ over six
tested values, resulting in 18 tested $(R_{sample},\beta)$
configurations.
For each configuration, 20 Monte Carlo trials are performed, and their
mean runtime and mean NMSE form one configuration-level operating point.
The empirical covariance ellipse in
Fig.~\ref{Fig_RuntimeNMSEEnvelope} is computed in the
$(\log_{10}(\mathrm{runtime}),\log_{10}(\mathrm{NMSE}))$ plane from these
18 configuration-level mean points.
Therefore, each ellipse characterizes how the average accuracy--runtime
operating point of a baseline varies with its codebook/search parameters,
rather than the trial-to-trial Monte Carlo variability.

The major-axis direction and extent of each ellipse characterize the
dominant joint variation of runtime and NMSE as $(R_{sample},\beta)$ changes,
while its tilt indicates the prevailing accuracy--runtime trade-off.
The minor-axis extent describes the residual configuration-dependent
spread around this dominant trend, and the ellipse area summarizes the
overall sensitivity of the operating point to the tested codebook
settings.
Accordingly, a more compact ellipse indicates greater robustness to
codebook/search-parameter selection, whereas a broader ellipse indicates
stronger configuration dependence.

The OMP- and P-SIGW-based ellipses exhibit pronounced negative-slope
major axes, showing that increasing the spatial search resolution
generally improves accuracy at the cost of additional computation.
The extremely fast operating points correspond to very coarse or small
codebooks, which substantially reduce the number of candidate
evaluations but also lead to sharp NMSE degradation.
Hence, these points represent the low-computation, low-accuracy end of
the corresponding accuracy--runtime trade-off rather than comparable
accuracy at substantially lower cost.
In contrast, the Joint-ML ellipse is more vertically oriented, indicating
that the tested configurations produce substantial runtime variation
with comparatively limited systematic NMSE improvement.

The proposed estimator does not depend on the baseline codebook
parameters $R_{sample}$ and $\beta$ and therefore has no corresponding
configuration ellipse.
Instead, the serial--dense and parallel--banded implementations are
represented by their mean runtime--NMSE operating points together with
Monte Carlo error bars, which characterize their trial-to-trial
variability.
Both proposed operating points lie on the lower-NMSE side of the tested
baseline operating regions at comparable runtime.
This indicates that, within the tested search range, the accuracy gain
of the proposed motion-native inference is not reproduced simply by
retuning the snapshot-first codebook resolution.

Finally, replacing the serial--dense backend with the
parallel--banded implementation yields a $1.86\times$ runtime reduction
with essentially unchanged NMSE.
This comparison isolates the implementation-level acceleration of the
same proposed estimator and should be distinguished from the
accuracy--runtime trade-off across different baseline configurations.
Overall, Fig.~\ref{Fig_RuntimeNMSEEnvelope} demonstrates a favorable
tested accuracy--runtime operating point without implying a universal
runtime advantage over all possible implementations or search
configurations.

\section{Conclusion}

In this paper, we developed a motion-native framework for continuous
channel tracking in THz near-field XL-MIMO systems by treating the UE
trajectory over each tracking window as the primary inference object.
A cubic B-spline representation couples sparse pilot observations
through one shared trajectory before any pilot-local state is finalized.
We characterized the
observation-supported identifiability of the shared trajectory and
showed that trajectory-relevant information lost during
parameter-independent snapshot-first processing cannot, in general, be
recovered by subsequent temporal processing. These results clarify when
complementary pilot observations can jointly constrain the trajectory
and why direct observation-level coupling can outperform
estimate-then-track or estimate-then-smooth pipelines.

To solve the resulting nonlinear problem efficiently, we introduced a
lifted local--global formulation with parallel pilot-local geometric
updates and a structured global spline update. The compact support of
the B-spline basis yields a fixed-bandwidth global system with
linear-complexity factorization for a fixed spline degree. Simulations
demonstrate improved tracking accuracy over representative snapshot-first
baselines across the tested SNR, propagation-range, observation-support,
smooth-motion, and dominant-path-mismatch settings. The results further
characterize regularization sensitivity, optimization behavior, and the
end-to-end accuracy--runtime trade-off enabled by the structured
implementation.

The present formulation focuses on a smooth-motion, dominant-trackable-path, single-user batch setting. Future work will extend the framework to wideband hybrid-field and multipath propagation, investigate physics-constrained deep unfolding of the structured solver, and develop joint multi-user tracking under practical synchronization, calibration, and interference impairments.

\appendices

\section{Proofs of the Identifiability Results}
\label{app:identifiability_proof}

\begin{IEEEproof}
	We first establish the general rank characterization \eqref{eq:weighted_sampling_rank} in
	Proposition~\ref{prop:weighted_sampling_identifiability}.
	By construction, each block of $\mathbf A_{\mathcal O}$ is the
	pilot-local geometric information weighted by its spline mapping.
	Therefore, stacking all pilot contributions gives the Gram
	representation
	\begin{equation}
		\mathcal I_{\boldsymbol\varpi}^{\rm obs}
		=
		\sum_{k=1}^{K}
		\mathbf S_k^{\mathsf T}
		\mathcal I_{\mathbf p,k}^{\rm obs}
		\mathbf S_k
		=
		\mathbf A_{\mathcal O}^{\mathsf T}
		\mathbf A_{\mathcal O}.
		\label{eq:app_information_gram}
	\end{equation}
	This representation directly connects observation-supported
	identifiability to the column space of
	$\mathbf A_{\mathcal O}$. In particular, for any nonzero
	$\delta\boldsymbol\varpi\in\mathbb R^{2L}$,
	\begin{equation}
		\delta\boldsymbol\varpi^{\mathsf T}
		\mathcal I_{\boldsymbol\varpi}^{\rm obs}
		\delta\boldsymbol\varpi
		=
		\left\|
		\mathbf A_{\mathcal O}\delta\boldsymbol\varpi
		\right\|_2^2 .
		\label{eq:app_information_quadratic}
	\end{equation}
	Hence,
	$\mathcal I_{\boldsymbol\varpi}^{\rm obs}\succ\mathbf0$
	holds if and only if
	$\mathbf A_{\mathcal O}$ has no nontrivial null direction,
	i.e.,
	$\operatorname{rank}(\mathbf A_{\mathcal O})=2L$. The same representation also yields the necessary
	information-dimension condition. If
	$\operatorname{rank}(\mathbf A_{\mathcal O})=2L$, then
	\begin{equation}
		\begin{aligned}
			2L
			&=
			\operatorname{rank}(\mathbf A_{\mathcal O})
			\\
			&\le
			\sum_{k=1}^{K}
			\operatorname{rank}
			\left(
			(\mathcal I_{\mathbf p,k}^{\rm obs})^{1/2}
			\mathbf S_k
			\right)
			\\
			&\le
			\min
			\left\{
			\sum_{k=1}^{K}
			\operatorname{rank}
			\left(
			(\mathcal I_{\mathbf p,k}^{\rm obs})^{1/2}
			\right),
			\sum_{k=1}^{K}
			\operatorname{rank}(\mathbf S_k)
			\right\}
			\\
			&=
			\min
			\left\{
			\sum_{k=1}^{K}
			\operatorname{rank}
			\left(
			\mathcal I_{\mathbf p,k}^{\rm obs}
			\right),
			2K
			\right\}\le 2K.
		\end{aligned}
		\label{eq:app_rank_necessary}
	\end{equation}
	Since $\mathcal I_{\mathbf p,k}^{\rm obs}\in\mathbb R^{2\times 2}$ and
	$\operatorname{rank}(\mathbf S_k)=2$, \eqref{eq:app_rank_necessary} implies both
	$\sum_{k=1}^{K}\operatorname{rank}\!\left(\mathcal I_{\mathbf p,k}^{\rm obs}\right)\ge 2L$
	and $K\ge L$.
	
	Beyond the preceding condition, we next establish the B-spline sampling condition in
	\eqref{eq:bspline_collocation_rank}. The necessary dimension condition
	$K\ge L$ alone does not guarantee identifiability, because the pilot
	sampling pattern must also span the complete B-spline coefficient space. From the definition of $\mathbf A_{\mathcal O}$,
	\begin{equation}
		\mathbf A_{\mathcal O}\delta\boldsymbol\varpi
		=
		\begin{bmatrix}
			(\mathcal I_{\mathbf p,1}^{\rm obs})^{1/2}
			\mathbf S_1\delta\boldsymbol\varpi\\
			\vdots\\
			(\mathcal I_{\mathbf p,K}^{\rm obs})^{1/2}
			\mathbf S_K\delta\boldsymbol\varpi
		\end{bmatrix}.
		\label{eq:app_weighted_perturbation}
	\end{equation}
	If the \(k\)th pilot has full local identifiability, i.e., \(\mathcal I_{\mathbf p,k}^{\rm obs}\succ0\) for every \(k\), then each \((\mathcal I_{\mathbf p,k}^{\rm obs})^{1/2}\) is nonsingular.
	Therefore, the information weighting does not change the
	pilot-wise null directions, and
	\begin{equation}
		\mathbf A_{\mathcal O}\delta\boldsymbol\varpi=\mathbf0
		\quad\Longleftrightarrow\quad
		\mathbf S_k\delta\boldsymbol\varpi=\mathbf0,\ \forall k.
		\label{eq:app_nullspace_equivalence}
	\end{equation}
	Since
	$\mathbf S_k
	=\mathbf b^{\mathsf T}(t_{n_k})\otimes\mathbf I_2$,
	the stacked spline constraints are equivalently represented by
	$\mathbf B_{\mathcal O}\otimes\mathbf I_2$. Therefore,
	\begin{equation}
		\operatorname{rank}(\mathbf A_{\mathcal O})
		=
		\operatorname{rank}
		(\mathbf B_{\mathcal O}\otimes\mathbf I_2)
		=
		2\operatorname{rank}(\mathbf B_{\mathcal O}).
		\label{eq:app_bspline_rank}
	\end{equation}
	Combining this relation with
	\eqref{eq:weighted_sampling_rank} gives
	\[
	\mathcal I_{\boldsymbol\varpi}^{\rm obs}\succ\mathbf0
	\quad\Longleftrightarrow\quad
	\operatorname{rank}(\mathbf B_{\mathcal O})=L,
	\]
	which proves~\eqref{eq:bspline_collocation_rank}. For a degree-$q$ B-spline, the latter full-column-rank
	condition is guaranteed if the pilot set contains $L$ ordered
	sites satisfying the Schoenberg--Whitney conditions
	$B_{i,q}(\widetilde s_i)>0$~\cite{de1978practical}.
	Indeed, the corresponding $L\times L$ row submatrix of
	$\mathbf B_{\mathcal O}$ is nonsingular, which implies
	$\operatorname{rank}(\mathbf B_{\mathcal O})=L$.
	This completes the proof.
\end{IEEEproof}

\section{Proof of Proposition~\ref{prop:information_monotonicity}}
\label{app:information_relation}

\begin{IEEEproof}
	Considering the complete snapshot-first processing chain $\widehat{\boldsymbol\xi}_k
	=\mathcal K_{{\rm s},k}(\mathbf y_k)$ in \eqref{eq:Markovian Chain}, both mappings are
	parameter independent under the standard regularity conditions. Thus, the data-processing inequality implies that a subsequent parameter-independent statistical mapping cannot increase efficient
	Fisher information as
	\begin{equation}
		\mathcal I_{\mathbf p,k}^{\rm snap}
		\preceq
		\mathcal I_{\mathbf p,k}^{\rm obs},
		\qquad k=1,\ldots,K .
		\label{eq:app_local_snapshot_dpi}
	\end{equation}
	Since
	$\mathbf p_k=\mathbf S_k\boldsymbol\varpi$, the chain rule maps the
	efficient information retained by the $k$th snapshot into the
	coefficient space as
	$\mathbf S_k^{\mathsf T}
	\mathcal I_{\mathbf p,k}^{\rm snap}\mathbf S_k$.
	Moreover, because the snapshot mappings remain pilot-wise and the
	pilot observations are conditionally independent, the information
	retained by the complete snapshot sequence with respect to
	$\boldsymbol\varpi$ is
	\begin{equation}
		\mathcal I_{\boldsymbol\varpi}^{\rm snap}
		\triangleq
		\sum_{k=1}^{K}
		\mathbf S_k^{\mathsf T}
		\mathcal I_{\mathbf p,k}^{\rm snap}
		\mathbf S_k .
		\label{eq:app_snapshot_information_pullback}
	\end{equation}
	Similarly, Proposition~\ref{prop:weighted_sampling_identifiability}
	gives the raw-observation information
	\begin{equation}
		\mathcal I_{\boldsymbol\varpi}^{\rm obs}
		=
		\sum_{k=1}^{K}
		\mathbf S_k^{\mathsf T}
		\mathcal I_{\mathbf p,k}^{\rm obs}
		\mathbf S_k .
		\label{eq:app_observation_information_pullback}
	\end{equation}
	Subtracting \eqref{eq:app_snapshot_information_pullback} from
	\eqref{eq:app_observation_information_pullback} and using
	\eqref{eq:app_local_snapshot_loss} yields
	\begin{equation}
		\mathcal I_{\boldsymbol\varpi}^{\rm obs}
		-
		\mathcal I_{\boldsymbol\varpi}^{\rm snap}
		=
		\sum_{k=1}^{K}
		\mathbf S_k^{\mathsf T}
		\Delta\mathcal I_{\mathbf p,k}^{\mathcal K_{\rm s}}
		\mathbf S_k
		\succeq\mathbf0 .
		\label{eq:app_snapshot_loss_propagation}
	\end{equation}
	Hence, information discarded during pilot-local snapshot formation is
	mapped into the shared coefficient space through the same spline
	operators that aggregate complementary pilot information in
	Proposition~\ref{prop:weighted_sampling_identifiability}.
	
	We then consider the subsequent temporal processing
	$\widehat{\boldsymbol\zeta}
	=\mathcal K_{\rm p}(\widehat{\boldsymbol\xi})$.
	Applying the same data-processing principle to this
	parameter-independent mapping, now with respect to the shared
	trajectory parameter $\boldsymbol\varpi$, gives
	\begin{equation}
		\mathcal I_{\boldsymbol\varpi}^{\rm post}
		\preceq
		\mathcal I_{\boldsymbol\varpi}^{\rm snap}.
		\label{eq:app_postprocessing_dpi}
	\end{equation}
	Therefore, the additional trajectory-information loss introduced by
	temporal postprocessing is
	\begin{equation}
		\Delta\mathcal I_{\boldsymbol\varpi}^{\mathcal K_{\rm p}}
		\triangleq
		\mathcal I_{\boldsymbol\varpi}^{\rm snap}
		-
		\mathcal I_{\boldsymbol\varpi}^{\rm post}
		\succeq\mathbf0 .
		\label{eq:app_postprocessing_loss}
	\end{equation}
	
	Finally, using
	$\Delta\mathcal I_{\boldsymbol\varpi}^{\rm tot}
	\triangleq
	\mathcal I_{\boldsymbol\varpi}^{\rm obs}
	-
	\mathcal I_{\boldsymbol\varpi}^{\rm post}$,
	we decompose
	\begin{equation}
		\begin{aligned}
			\Delta\mathcal I_{\boldsymbol\varpi}^{\rm tot}
			&=
			\left(
			\mathcal I_{\boldsymbol\varpi}^{\rm obs}
			-
			\mathcal I_{\boldsymbol\varpi}^{\rm snap}
			\right)
			+
			\left(
			\mathcal I_{\boldsymbol\varpi}^{\rm snap}
			-
			\mathcal I_{\boldsymbol\varpi}^{\rm post}
			\right)
			\\
			&=
			\sum_{k=1}^{K}
			\mathbf S_k^{\mathsf T}
			\Delta\mathcal I_{\mathbf p,k}^{\mathcal K_{\rm s}}
			\mathbf S_k
			+
			\Delta\mathcal I_{\boldsymbol\varpi}^{\mathcal K_{\rm p}}
			\succeq\mathbf0 ,
		\end{aligned}
		\label{eq:app_total_information_loss}
	\end{equation}
	which proves~\eqref{eq:total_information_loss}. It remains to establish the zero-loss condition in
	\eqref{eq:zero_information_loss}. 
	
	If
	$\Delta\mathcal I_{\mathbf p,k}^{\mathcal K_{\rm s}}=\mathbf0$
	for every $k$ and
	$\Delta\mathcal I_{\boldsymbol\varpi}^{\mathcal K_{\rm p}}
	=\mathbf0$, then
	\eqref{eq:app_total_information_loss} immediately gives
	$\Delta\mathcal I_{\boldsymbol\varpi}^{\rm tot}=\mathbf0$. Conversely, suppose that
	$\Delta\mathcal I_{\boldsymbol\varpi}^{\rm tot}=\mathbf0$.
	All terms on the right-hand side of
	\eqref{eq:app_total_information_loss} are positive semidefinite, and
	therefore none can cancel another. It follows that
	\begin{equation}
		\Delta\mathcal I_{\boldsymbol\varpi}^{\mathcal K_{\rm p}}
		=\mathbf0,
		\qquad
		\mathbf S_k^{\mathsf T}
		\Delta\mathcal I_{\mathbf p,k}^{\mathcal K_{\rm s}}
		\mathbf S_k
		=\mathbf0,
		\quad \forall k .
		\label{eq:app_zero_loss_components}
	\end{equation}
	For the adopted B-spline representation,
	$\mathbf S_k
	=\mathbf b^{\mathsf T}(t_{n_k})\otimes\mathbf I_2$.
	Since the B-spline basis forms a partition of unity,
	$\mathbf b(t_{n_k})\neq\mathbf0$ at every valid pilot instant, and
	hence
	$\operatorname{rank}(\mathbf S_k)=2$.
	Thus, $\mathbf S_k$ has full row rank, yielding
	$\Delta\mathcal I_{\mathbf p,k}^{\mathcal K_{\rm s}}
	=\mathbf0$ for all $k$.
	Together with
	$\Delta\mathcal I_{\boldsymbol\varpi}^{\mathcal K_{\rm p}}
	=\mathbf0$, this proves
	\eqref{eq:zero_information_loss}.
\end{IEEEproof}

\section{Proofs of the B-Spline Structural Properties}
\label{app:bspline_structural_proofs}

\begin{IEEEproof}[Proof of Lemma~\ref{lem:bspline_local_support}]
	For $q=0$, the definition of the degree-zero basis gives $	\operatorname{supp}\!\left(B_{\ell,0}\right)
	\subseteq
	[\tau_\ell,\tau_{\ell+1}]$. Assume that
	\[
	\operatorname{supp}\!\left(B_{\ell,q-1}\right)
	\subseteq
	[\tau_\ell,\tau_{\ell+q}].
	\]
	By the Cox--de Boor recursion in \eqref{eq:cox_de_boor},
	$B_{\ell,q}$ is a weighted sum of
	$B_{\ell,q-1}$ and $B_{\ell+1,q-1}$. Since the scalar
	weights do not enlarge their supports,
	\[
	\begin{aligned}
		\operatorname{supp}\!\left(B_{\ell,q}\right)
		&\subseteq
		\operatorname{supp}\!\left(B_{\ell,q-1}\right)
		\cup
		\operatorname{supp}\!\left(B_{\ell+1,q-1}\right)\\
		&\subseteq
		[\tau_\ell,\tau_{\ell+q+1}].
	\end{aligned}
	\]
	The support bound therefore follows by induction. If
	$s\in[\tau_j,\tau_{j+1})$, only the consecutive indices
	$\ell\in\{j-q,\ldots,j\}$ can be active. Hence, at most
	$q+1$ basis functions are nonzero at any $s\in[0,1]$,
	including the right endpoint under the adopted clamped-knot
	convention.
\end{IEEEproof}

\begin{IEEEproof}[Proof of Lemma~\ref{lem:global_banded_sparsity}]
	By Lemma~\ref{lem:bspline_local_support}, at most $q+1$
	consecutive B-spline basis functions are active at any pilot instant.
	Consequently, $\mathbf S_k$ contains at most $q+1$ consecutive
	active control-point blocks, and
	$[\mathbf S_k^{\mathsf T}\mathbf S_k]_{\ell,r}
	=\mathbf0_{2\times2}$ whenever $|\ell-r|>q$.
	
	Moreover, $\mathbf D_1$ and $\mathbf D_2$ are the first- and
	second-order difference operators, so
	$\mathbf D_1^{\mathsf T}\mathbf D_1$ and
	$\mathbf D_2^{\mathsf T}\mathbf D_2$ have block half-bandwidths at
	most one and two, respectively. Hence, the regularization matrix
	$\mathbf R$ has block half-bandwidth at most two. Since
	$\mathbf C_{\boldsymbol\varpi}
	=\mathbf R+\eta\sum_{k=1}^{K}
	\mathbf S_k^{\mathsf T}\mathbf S_k$, it follows that
	\[
	[\mathbf C_{\boldsymbol\varpi}]_{\ell,r}
	=
	\mathbf0_{2\times2},
	\qquad
	|\ell-r|>\max\{q,2\}.
	\]
	Therefore, \(\mathbf C_{\boldsymbol\varpi}\) is symmetric and block banded with block half-bandwidth at most \(b_C=\max\{q,2\}\), independently of \(L\).
\end{IEEEproof}

\bibliographystyle{IEEEtran}
\bibliography{references}

\newpage
\begin{IEEEbiography}[{\includegraphics[width=1in,height=1.25in,clip,keepaspectratio]{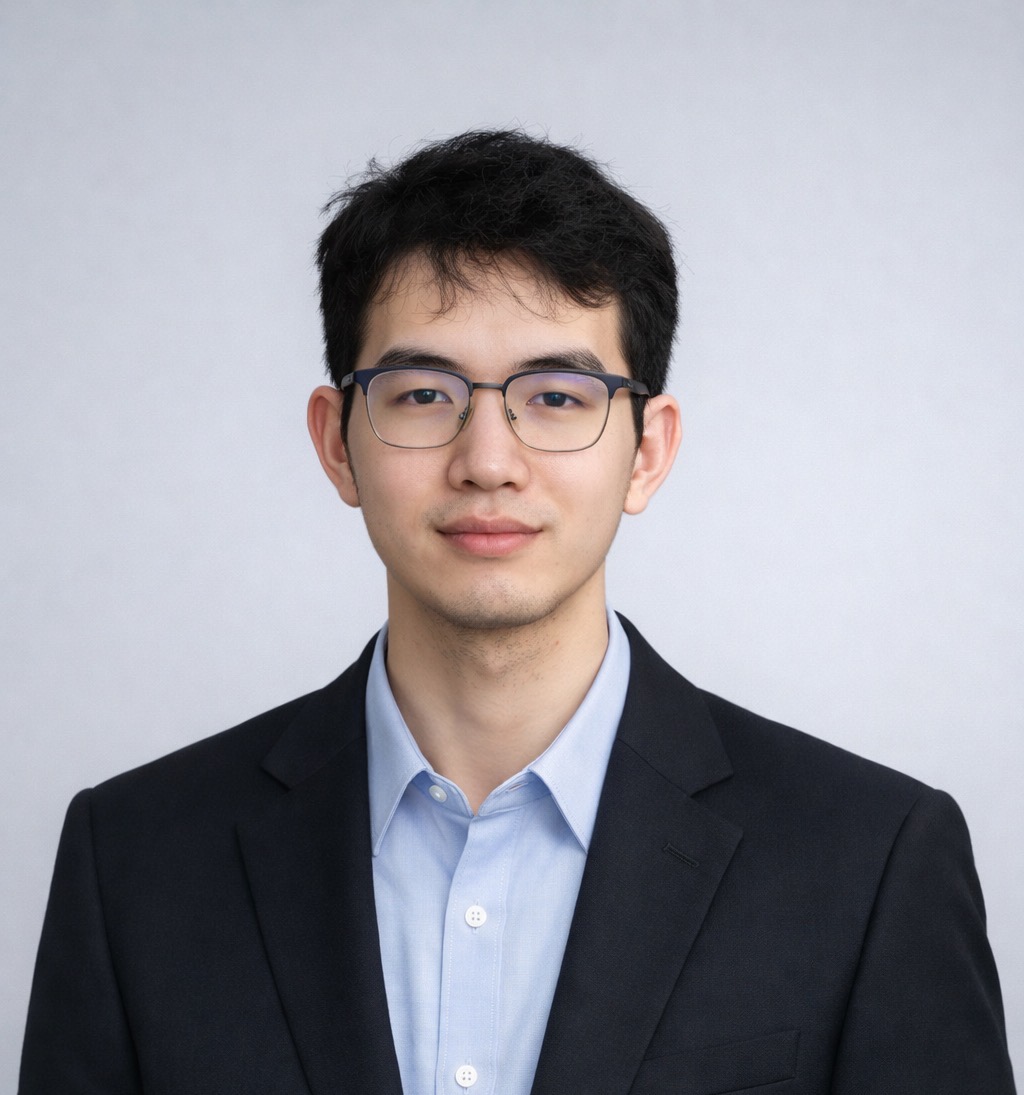}}]{Yuheng Fan}
received the B.S. degree in information and communication engineering from Beijing University of Posts and Telecommunications, Beijing, China, in 2023, where he is currently pursuing the Ph.D. degree in information and communication engineering with the State Key Laboratory of Networking and Switching Technology. His research interests include signal processing for massive MIMO terahertz communication systems.
\end{IEEEbiography}

\begin{IEEEbiography}[{\includegraphics[width=1in,height=1.25in,clip,keepaspectratio]{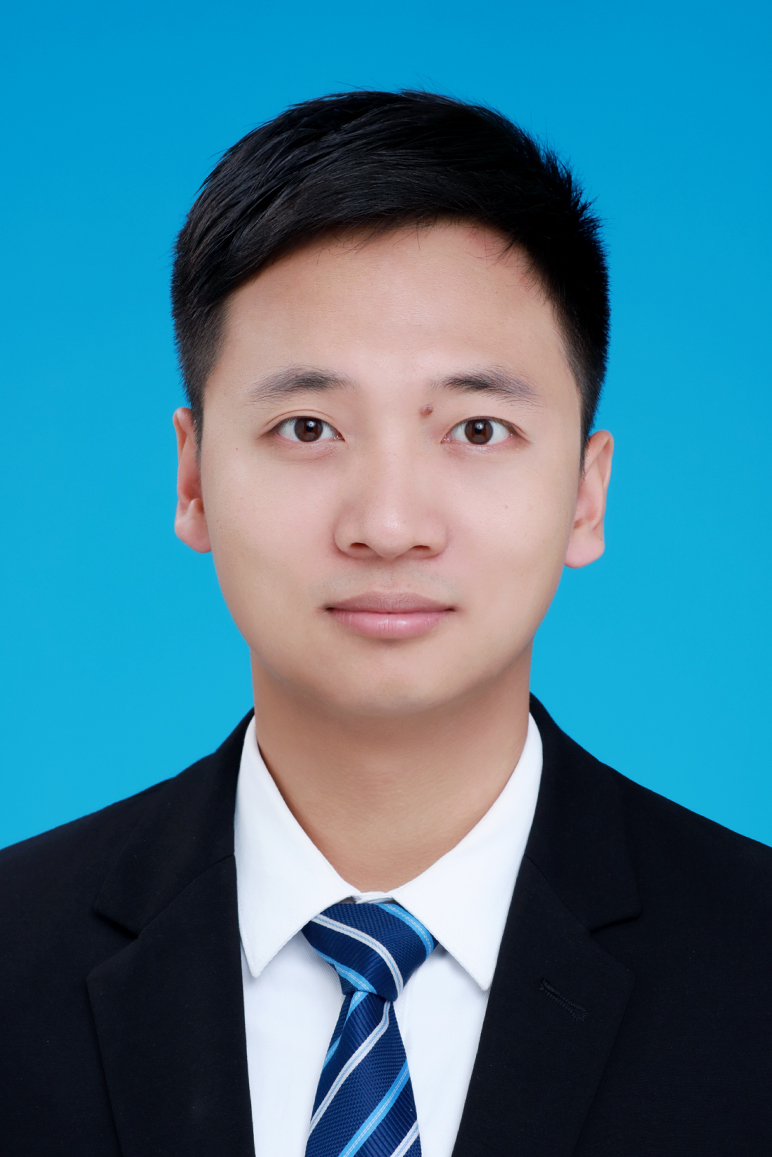}}]{Chuang Yang (Member, IEEE)}
received the B.S. degree in electronic information science and technology from Chongqing University, Chongqing, China, in 2015, and the Ph.D. degree in microelectronics and solid-state electronics from Tianjin University, Tianjin, China, in 2020.
He is currently an Associate Researcher with the State Key Laboratory of Networking and Switching Technology, Beijing University of Posts and Telecommunications, Beijing, China. His research interests include microwave and terahertz measurements, and terahertz integrated sensing and communication.
\end{IEEEbiography}

\begin{IEEEbiography}[{\includegraphics[width=1in,height=1.25in,clip,keepaspectratio]{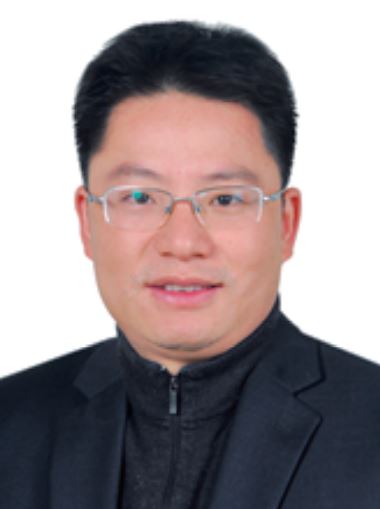}}]{Mugen Peng (Fellow, IEEE)}
received the Ph.D. degree in communication and information systems from Beijing University of Posts and Telecommunications (BUPT), Beijing, China, in 2005.
Afterward, he joined BUPT, where he has been a Full Professor with the School of Information and Communication Engineering since 2012. In 2014, he was also an Academic Visiting Fellow with Princeton University, Princeton, NJ, USA. He leads a research group focusing on wireless transmission and networking technologies at BUPT. He has authored and coauthored over 100 refereed IEEE journal papers and over 300 conference papers. His main research areas include wireless communication theory, radio signal processing, cooperative communication, self-organization networking, heterogeneous networking, cloud communication, and the Internet of Things.
Prof. Peng was a recipient of the 2018 Heinrich Hertz Prize Paper Award; the 2014 IEEE ComSoc AP Outstanding Young Researcher Award; and Best Paper Awards from IEEE ICC 2022, JCN 2016, IEEE WCNC 2015, IEEE GameNets 2014, IEEE CIT 2014, ICCTA 2011, IC-BNMT 2010, and IET CCWMC 2009. He is on the editorial or associate editorial boards of IEEE Communications Magazine, IEEE Access, IEEE Internet of Things Journal, IET Communications, and China Communications. He is a Fellow of the IET.
\end{IEEEbiography}

\vfill

\end{document}